\documentclass{aa}  
\usepackage{graphicx}
\usepackage{txfonts}
\newcommand{\lsim}{\raisebox{-5pt}{$\;\stackrel{\textstyle <}{\sim}\;$}}
\newcommand{\gsim}{\raisebox{-5pt}{$\;\stackrel{\textstyle >}{\sim}\;$}}

\begin{document}
   \title{The color excess of quasars with intervening DLA systems}

   \subtitle{Analysis of the SDSS data release five}

   \author{  Giovanni Vladilo\inst{1}, 
            Jason X. Prochaska\inst{2},  Arthur M. Wolfe\inst{3} 
          }

\titlerunning{color excess of QSOs with intervening DLA systems}         
\authorrunning{Vladilo,  Prochaska, \& Wolfe } 

   \offprints{G. Vladilo}

   \institute{ 
Osservatorio Astronomico di Trieste, Istituto Nazionale di Astrofisica,  Trieste, Italy \\
              \email{vladilo@oats.inaf.it}
         \and
Department of Astronomy and Astrophysics, UCO/Lick Observatory, University of California, Santa Cruz, CA, USA\\
             \email{xavier@ucolick.org}
         \and
Department of Physics, and Center for Astrophysics and Space Sciences,  
University of California at San Diego, La Jolla, CA, USA\\
             \email{awolfe@awolfe@.ucsd.edu}         
             }

   \date{Received ...; accepted ...}
  
\abstract{ 
    We analyzed the spectroscopic and photometric database 
    of the 5th data release of the Sloan Digital Sky Survey (SDSS)
   to search for evidence of the quasar reddening produced
   by dust embedded in intervening damped Ly\,$\alpha$ (DLA) systems.   
   From a list of 5164 quasars in the interval of emission redshift 
   $2.25 \leq z_e \leq 3.5$ and SDSS spectra with signal-to-noise ratio SNR $\gsim 4$,
   we built up an ``absorption sample'' of 248 QSOs with a single
   DLA system in the interval of absorption redshift  $2.2 < z_a \leq 3.5$
   and a  ``pool'' of $1959$ control QSOs without  DLA systems or strong  metal systems.              For each  QSO of the absorption sample we extracted from the pool 
    a subset of control QSOs that are closest in redshift and magnitude.
   The mean color of this subset was used as a zero point to measure the ``deviation 
   from the mean color'' of individual DLA-QSOs, $\Delta_i$. 
   The colors were measured using ``BEST'' $ugriz$ SDSS imaging data.
   The mean color excess of the absorption sample, $\langle E \rangle$, was estimated
   by averaging the individual color deviations $\Delta_i$.    We find  
        $\langle E(r-z) \rangle = 27 \, (\pm 9) \times 10^{-3}$ mag 
     and  ${\langle E(g-z) \rangle}  = 54 \, (\pm 12) \times 10^{-3}$ mag. These values are
     representative of the reddening of DLA systems at  $z_a \approx 2.7$
     in SDSS QSOs with limiting magnitude $r \simeq 20.2$.
  The detection of the mean reddening is confirmed by several statistical tests.  
    Analysis of the results suggests an origin  of the reddening   
    in dust embedded in the DLA systems, with an SMC-type extinction curve. 
    By converting the reddening into rest-frame extinction, we derive a mean dust-to-gas ratio 
    $\langle A_V / N(\ion{H}{i}) \rangle \approx 2$  to $4 \times 10^{-23}$ mag cm$^2$.
    This value is $\simeq -1.25$ dex lower than the mean dust-to-gas ratio of the Milky Way,
    in line with the lower level of metallicity in the present DLA sample. 
       }
\maketitle 
   
\section{Introduction}

Quasar absorption-line systems allow us to probe the physical properties  
 of intergalactic/interstellar gas over a large fraction
of the Hubble time, back  to the epoch of quasar formation. 
The quasar absorbers with
neutral hydrogen column density $N(\ion{H}{i}) \geq 10^{20.3}$ atoms cm$^{-2}$
  are believed to originate in the interstellar gas of intervening galaxies
(Wolfe, Gawiser, \& Prochaska 2005), rather than in the
intergalactic medium. 
The presence of a strong Ly\,$\alpha$ profile with ``damping'' wings extending well beyond
the Doppler core of the line is a distinctive feature that gives the
name to this class of absorbers (Wolfe et al. 1986).
Spectroscopic studies of ``damped Ly\,$\alpha$'' (DLA) systems 
yield  unique and accurate information on
gas-rich, high-redshift galaxies observed in absorption 
(hereafter   ``DLA galaxies'').

At redshift $z \lsim 1.6$  about 20
DLA host galaxies  have been imaged in the quasar field,
showing a variety of morphological types 
(Le Brun et al. 1997, Rao et al. 2003, Chen et al. 2004).
At higher redshift, where the bulk of known DLA systems are located,
the imaging identification of the galaxy is quite difficult and
most of the studies on the nature of DLA galaxies are based on spectroscopic observations. 
In spite of a long series of   studies on their
kinematics
(Prochaska \& Wolfe 1997, 1998), 
chemical abundances 
(Pettini et al. 1994; Lu et al. 1996;  Prochaska \& Wolfe 1999; 
Molaro et al. 2000; 
Dessauges-Zavadsky et al. 2006),
and physical properties 
(Petitjean et al. 2000; Wolfe, Prochaska \& Gawiser 2003),
the exact nature of DLA galaxies is still open to debate (Wolfe \& Chen 2006).
Studies of DLA systems allow us to probe faint galaxies that happen to lie
in the direction of the quasar.
It is not clear whether there is a continuity or not between the 
properties of absorption-selected DLA galaxies   and 
 photometrically selected Ly-break galaxies, which are biased in luminosity 
(e.g. M\o ller et al. 2004).

In the present work we focus our attention on the dust
component of DLA systems.
By analogy with studies of local galaxies, we expect interstellar dust
to be a pervasive component of DLA galaxies.
The properties of dust are poorly understood, but may play a critical role 
in several investigations aimed at  understanding the nature of DLA galaxies.  
In fact, dust is expected to play a major role in a variety of
physical  processes 
and to affect the measurement of observational quantities.

An example of the importance of dust in governing {\em the physical processes} at work
in DLA galaxies comes from the study of the \ion{C}{ii}$^\ast$ 1335.7\,\AA\ line, 
which can be used to estimate the star formation rate assuming that the heating is dominated
by photoelectric emission from dust grains 
(Wolfe, Prochaska \& Gawiser 2003).
The efficiency of the heating mechanism and the accuracy
of the SFR determined in this way depend on the dust-to-gas ratio, 
which is poorly constrained. 
Also the production of molecular gas is believed to be influenced 
by the presence of dust, which acts as a catalyst of H$_2$ formation. 

An example of the impact of dust on {\em the observations} of DLA systems 
comes from the different estimates of chemical abundances obtained from
refractory   and volatile elements (Pettini et al. 1994; 
Vladilo 1998, 2002; Centuri\'on et al. 2000; Hou et al. 2001, Prochaska \& Wolfe 2002). 
A significant fraction of refractory elements is depleted from the gas phase,
where they can be detected via absorption line spectroscopy, to the dust component,
where they escape detection with this technique.
Studies of galactic chemical evolution are based on measurements
of abundance ratios which may be affected by differential dust depletion.
The inferences of galactic chemical evolution models
on the  nature of DLA galaxies depend on
the exact amount  of dust depletion in these systems
(e.g. Calura et al. 2003, Dessauges-Zavadsky et al. 2006). 

Another observational effect that we expect to occur
is the absorption and scattering of the photons of the quasar continuum
by dust embedded in the intervening DLA system. 
This extinction process is more efficient at shorter wavelengths
and is expected to redden the quasar colors. 
In the most extreme cases the
extinction could lead to the obscuration of the
background quasars  
(Ostriker \& Heisler 1984; Fall \& Pei 1989, 1993).
In the more general case, the extinction may induce a 
selection bias acting against the detection of metal-rich galactic regions in magnitude-limited
surveys of quasars (Prantzos \& Boissier 2000, Vladilo \& P\`eroux 2005).
Studies of radio-selected quasars surveys suggest that the impact  of this effect 
on the statistical properties of DLA systems is modest
(Ellison et al. 2001, Akerman et al. 2005, Jorgenson et al. 2006), 
but the size of these surveys is not sufficiently large to reach firm conclusions. 
The existence of empty fields without optical identifications
in the radio-selected survey of  Jorgenson et al. (2006)
suggests that a fraction of quasars may indeed be obscured. 
A better  understanding of the amount and type
of dust present in DLA systems is fundamental to
assess the importance of the dust extinction effect.

Because of the potential impact of dust in studies of DLA systems,
a significant observational effort has been dedicated to
prove its existence and to understand its properties. 
At  redshift $z \lsim 2$  definitive evidence for DLA dust  
has been found in one line of sight
where the dust extinction bump at 2175\,\AA\
and the silicate  absorption at 9.7\,$\mu$ have been detected 
(Junkkarinen et al. 2004, Kulkarni et al. 2007). 
Evidence for reddening due to DLA absorbers at  redshift $z \lsim 2$ 
has been found in a few quasars with
metal rich DLA systems (Vladilo et al. 2006).
At $z \gsim 2$ the evidence for dust in DLA systems 
is mostly based  on studies of elemental depletions 
(Pettini et al. 1997, 2000; Hou et al. 2001; Prochaska \& Wolfe 2002; Vladilo 1998, 2004;  
Dessauges-Zavadsky et al. 2006) and on the existence of general
trends between depletion, metallicity and H$_2$ molecular fraction
(Ledoux et al. 2003, Petitjean et al. 2006).
In order to establish more firmly the presence of dust at $z \gsim 2$
and to understand its properties, it is fundamental to complement the
studies of depletion with measurements of quasar reddening.
A claim of reddening detection was reported by Pei et al. (1991),
based on the analysis of a sample of 13 DLA-quasars.
%
This claim was not confirmed by subsequent work, including
a photometric study of the colors of radio-selected quasars with and
without DLA systems (Ellison et al. 2005)
and a study of  quasar spectra of the 2nd data release of the Sloan Digital Sky Survey (SDSS) 
(Murphy \& Liske 2004).
The  remarkably low  upper limit of reddening found by Murphy \& Liske, 
 $\langle E(B-V) \rangle < 0.01$ mag
in the absorber frame,   indicates how challeging this type of measurement is. 
On the other hand, the detection of
quasar reddening due to \ion{Ca}{ii} systems
(Wild \& Hewett 2005, Wild et al. 2006)   
and \ion{Mg}{ii}   systems
(York et al. 2006, Menard et al. 2007)
suggests that also DLA systems should produce a reddening signal.

The aim of the present work is to exploit the  
large database of the 5th SDSS Data Release, in conjunction with novel
features in the analysis, 
to   detect   the mean reddening of DLA-QSOs at $z > 2$. 
As in previous work we use the   spectroscopic database in the process of selection of
the quasars of the absorption and control samples.
At variance with previous work  on DLA reddening, 
the  quasar colors are measured making use
of the photometric database. 
The  photometric measurement of the reddening 
avoids some critical steps inherent to the spectroscopic method,
such as the photometric calibration of the spectra and the tracing of the quasar continuum. 
Uncertainties in the photometric calibration, even if small (Adelman-McCarthy et al. 2007), translate into uncertainties in the slope of the continuum.

In  Section 2 we describe
the process of selection of the quasar samples.
Novel features of our approach include (i)  the use of spectra with the same signal-to-noise ratio
for the selection of DLA quasars and control quasars, (ii) the rejection of  
low-redshift absorption systems that may affect the measurement of the reddening  at high-redshift, and
(iii) the rejection of quasars with multiple DLA systems.
Thanks to these last criteria we are in the position
to search for correlations between the color excess and
the properties of individual DLA systems.
In Section 3 we explain the method adopted to measure the
mean  reddening of the absorption sample.
The measurement is presented in Section 4 and 
its interpretation is discussed in Section 5.

\section{The quasar samples}

The starting list of quasars was extracted with the SDSS DR5 spectroscopic query format.
A total of 7294 objects with spectroscopic constraints ``QSO'' or ``HiZ\_QSO'', 
emission redshift $2.25 \leq z_e \leq 3.5$ and imaging constraints
``Point Sources'' were extracted  from the query.
Quasars at lower redshift do not yield sufficient redshift path 
to search for DLA absorptions.
The  upper limit on $z_e$ comes from the requirement
that the photometric band $r$ 
falls redwards of the quasar Ly $\alpha$ emission (see Section 3.2).

The detectability of the   spectral signatures suggestive of reddening
  depends  on the signal-to-noise ratio (SNR) of their SDSS spectra.  
{\em In order to perform a homogeneous comparison between absorption and control samples
  we defined  a common threshold value of SNR. }
In practice, we adopted  the criterion
SNR $\geq$ 4, where SNR is the mean
signal-to-noise ratio  per pixel calculated in the spectral window 1440-1490\AA\ in
the rest frame of the quasar. 
We found  5164 quasars with $2.25 \leq z_e \leq 3.5$ and
SNR $\geq$ 4 in the adopted spectral window.


From this  list of quasars we extracted
the ``absorption sample'' and the ``control sample''.
The absorption sample is the subset of
{\em   
quasars with a single DLA system and without  any other
potential sources of foreground reddening.} 
The control sample is the subset of
{\em 
quasars without DLA systems or any other
potential sources of foreground reddening.} 

The identification of DLA systems and potential sources of reddening is based on
the analysis of   the   quasar spectra, as explained below.
 Concerning the DLA identification,
we adopt tight criteria in order  to {\em select} only 
genuine DLA systems. 
Concerning the other sources of reddening,
we instead adopt   broader criteria
since   we want  to be sure to
{\em reject}  any  potential  
source that may affect the reddening measurement.

The list of
signatures of potential sources of reddening 
that we considered includes: 
strong HI absorptions at redshift $z \neq z_\mathrm{DLA}$,
strong, low-ionization metal systems at $z \neq z_\mathrm{DLA}$, Broad Absorption Line (BAL) features, 
and strong self-absorptions in the quasar Ly  $\alpha$ emission. 
Strong metal lines of low ionization allow us to search for 
low-redshift intervening systems  
not directly detectable as a Ly  $\alpha$ line
because of  the limited spectral coverage of SDSS spectra. For instance, 
MgII absorbers are detectable down to redshift $z=0.95$ in quasars at $z_e=3.5$,
and down to $z=0.4$ in quasar at $z_e=2.25$. 
Other low-ionization metal lines redwards of the Ly  $\alpha$ emission 
(e.g. \ion{Si}{ii} 1526\,\AA, \ion{Fe}{ii} 2600\,\AA)
can trace potential sources of reddening at low redshift.
BAL features and strong Ly  $\alpha$ self-absorptions
arise  in the quasar environment and represent 
a reason of concern in the present analysis since they may be associated with dust. In fact,
BAL quasars tend to show peculiar colors relative to non-BAL ones (see e.g. Trump et al. 2006),
possibly as the result of dust present in the broad line region.
Strong HI self-absorptions at $z=z_e$ are suggestive of large amount of neutral gas, potentially
associated with dust.

\subsection{The selection process}

To build up the absorption sample we started by identifying  
Damped Ly\,$\alpha$ systems with the automated
algorithm described by Prochaska et al. (2005; hereafter PHW05)  
applied here to the
DR5.
Only systems with Ly\,$\alpha$ absorption blueshifted by at least 3000 km s$^{-1}$ relative to the quasar  Ly\,$\alpha$ emission\footnote
{
This  criterion   
  minimizes the likelihood that
the DLA absorption is physically associated with the quasar.
A study of the DLA systems closer than 3000 km s$^{-1}$ to the    Ly $\alpha$ emission
is presented in a separate work (Prochaska, Hennawi \& Herbert-Fort 2007).
}
were considered.
Quasars with BAL flags=1 and 2 according to the criteria explained in  PHW05
were rejected.  In this way
we obtained a list\footnote
{
The complete list of DLA systems of the DR5
including quasars with $z_e > 3.5$   an be found
at the web page
http://www.ucolick.org/$\sim$xavier/SDSSDLA.}
of 422 DLA systems in quasars with  $z_e \leq 3.5$
and SNR $\geq$ 4 in the rest-frame window 1440-1490\AA.
 
Starting from this list   the final absorption sample was built up
in the following way. 
We first discarded 69 quasars with strong MgII absorbers
at redshift $z_\mathrm{MgII} \neq z_\mathrm{DLA}$. In practice, 
due to the limitation of the SDSS wavelength coverage, we were able to search
for absorbers with $z_\mathrm{MgII} \lsim 2.2$.
To implement this rejection process we used the SDSS catalog of 
strong MgII absorbers 
with rest-frame equivalent width $\geq 1$ \AA\
 (Prochter et al. 2006) 
updated to DR5 (Prochter 2007, priv. comm.).
We then discarded 54 quasars with 
multiple DLA systems. 
Finally we rejected 30 quasars with a strong HI absorption at $z \neq z_\mathrm{DLA}$. 
For this purpouse we used a list of candidate HI absorptions
automatically generated.
We rejected cases with $N(\mathrm{HI}) \geq 10^{20}$ atoms cm$^{-2}$ within the errors.

After this automatic   process 
we visually inspected the   remaining 269  quasar spectra  to search for
spectroscopic features undetected by the automated algorithms.
A dozen of quasars with strong \ion{Mg}{ii} 
absorption  not associated with the DLA systems
were discarded in this way, as well as a few cases with anomalous spectra
(e.g. extremely narrow Ly  $\alpha$ emission). 
Other unidentified weak features 
were tentatively attributed to metal lines associated with the DLA system.

At the end of the rejection process we obtained a list of 248 DLA-quasar pairs
representing our absorption sample, listed in Table 1.
At the mean emission redshift  of this sample, $z_e \simeq 3$,
the  quasars are free of   foreground (MgII) systems
down to absorption redshift $z_a \sim 0.75$. 
The chance of intersecting an additional DLA system at lower redshifts
is low, so that the resulting DLA-quasar list is a good approximation of
an ideal sample of quasars with a single intervening DLA system at
high redshift ($2.2 \lsim z_a \lsim 3.5$).

 \begin{table*}[htdp]
\caption{
 Data for the quasars of the absorption sample and 
measurements of the deviations from the mean color 
$\Delta(r-z)$ and $\Delta(g-z)$.
Measurements in the $(g-z)$ color index are only given for quasars with $z_e \leq 3.4$ 
(see Section 3.2.2).
The first 20 DLA-quasars are shown. 
The full table is available  in electronic form
(see also web page http://adlibitum.oats.inaf.it/vladilo/tables.html).}
\begin{center}
\begin{tabular}{lccccccccrr}
\hline
SDSS & $z_e$ & $m_r$ & $z_a$ & $\log N(\ion{H}{i})$ & $\delta(z_e)^a$ & $\delta(m_z)^b$ & 
${\langle (r-z)_0 \rangle}^c$  & ${\langle (g-z)_0 \rangle}^c$ &  
$\Delta(r-z)^d$~~~~ & $\Delta(g-z)^d$~~~~ \\
 & & [mag] & & [cm$^{-2}$] & & [mag]~~ & [mag]~~~~ & [mag]~~~~ & [mag]~~~~ &  [mag]~~~~ \\
\hline
J0035-0918 & 2.42 & 18.89 & 2.338 & 20.55 & 0.03 & 0.26 & $ 0.229 \pm 0.101$ & $ 0.266 \pm 0.148$ & $ 0.191 \pm 0.111$ & $ 0.213 \pm 0.159$ \\ 
J0122+1334 & 3.01 & 19.16 & 2.349 & 20.30 & 0.09 & 0.36 & $ 0.109 \pm 0.175$ & $ 0.273 \pm 0.195$ & $ 0.010 \pm 0.189$ & $ 0.016 \pm 0.212$ \\ 
J0139-0824 & 3.02 & 18.44 & 2.677 & 20.70 & 0.09 & 0.54 & $ 0.111 \pm 0.133$ & $ 0.264 \pm 0.190$ & $-0.101 \pm 0.142$ & $-0.203 \pm 0.200$ \\ 
J0234-0751 & 2.54 & 18.84 & 2.319 & 20.95 & 0.05 & 0.23 & $ 0.200 \pm 0.117$ & $ 0.210 \pm 0.120$ & $-0.105 \pm 0.130$ & $-0.114 \pm 0.141$ \\ 
J0255-0711 & 2.82 & 19.09 & 2.612 & 20.45 & 0.13 & 0.37 & $ 0.150 \pm 0.173$ & $ 0.254 \pm 0.195$ & $-0.134 \pm 0.191$ & $-0.078 \pm 0.222$ \\ 
J0338-0005 & 3.05 & 18.20 & 2.229 & 20.90 & 0.10 & 1.70 & $ 0.170 \pm 0.164$ & $ 0.364 \pm 0.190$ & $-0.111 \pm 0.177$ & $-0.122 \pm 0.217$ \\ 
J0755+4056 & 2.35 & 19.04 & 2.301 & 20.35 & 0.03 & 0.26 & $ 0.208 \pm 0.117$ & $ 0.208 \pm 0.141$ & $-0.021 \pm 0.131$ & $ 0.034 \pm 0.160$ \\ 
J0840+5255 & 3.09 & 19.06 & 2.862 & 20.30 & 0.09 & 0.29 & $ 0.063 \pm 0.110$ & $ 0.260 \pm 0.129$ & $ 0.008 \pm 0.129$ & $-0.067 \pm 0.151$ \\ 
J0844+5153 & 3.20 & 19.16 & 2.775 & 21.45 & 0.10 & 0.31 & $ 0.106 \pm 0.109$ & $ 0.361 \pm 0.190$ & $-0.046 \pm 0.127$ & $-0.060 \pm 0.213$ \\ 
J0912+5621 & 3.00 & 18.94 & 2.889 & 20.55 & 0.09 & 0.50 & $ 0.116 \pm 0.131$ & $ 0.273 \pm 0.196$ & $ 0.314 \pm 0.137$ & $ 0.510 \pm 0.205$ \\ 
J0917+5917 & 2.40 & 18.98 & 2.329 & 20.35 & 0.03 & 0.21 & $ 0.214 \pm 0.101$ & $ 0.206 \pm 0.110$ & $-0.096 \pm 0.115$ & $ 0.234 \pm 0.127$ \\ 
J0919+5512 & 2.51 & 18.81 & 2.387 & 20.40 & 0.04 & 0.27 & $ 0.199 \pm 0.101$ & $ 0.240 \pm 0.125$ & $-0.075 \pm 0.109$ & $ 0.084 \pm 0.135$ \\ 
J0940+0232 & 3.21 & 19.18 & 2.565 & 20.70 & 0.10 & 0.36 & $ 0.075 \pm 0.086$ & $ 0.328 \pm 0.135$ & $-0.171 \pm 0.114$ & $-0.239 \pm 0.158$ \\ 
J1042+0117 & 2.44 & 18.31 & 2.267 & 20.75 & 0.03 & 0.41 & $ 0.203 \pm 0.128$ & $ 0.253 \pm 0.148$ & $-0.195 \pm 0.137$ & $-0.163 \pm 0.161$ \\ 
J1142-0012 & 2.49 & 18.58 & 2.258 & 20.35 & 0.04 & 0.38 & $ 0.240 \pm 0.136$ & $ 0.314 \pm 0.168$ & $ 0.109 \pm 0.141$ & $-0.088 \pm 0.175$ \\ 
J1208+0043 & 2.72 & 18.72 & 2.608 & 20.45 & 0.12 & 0.37 & $ 0.172 \pm 0.133$ & $ 0.288 \pm 0.179$ & $-0.040 \pm 0.138$ & $-0.025 \pm 0.185$ \\ 
J1228-0104 & 2.65 & 18.16 & 2.263 & 20.40 & 0.08 & 1.59 & $ 0.236 \pm 0.156$ & $ 0.273 \pm 0.143$ & $ 0.417 \pm 0.163$ & $ 0.437 \pm 0.152$ \\ 
J1251+6616 & 3.02 & 19.25 & 2.777 & 20.45 & 0.09 & 0.52 & $ 0.113 \pm 0.130$ & $ 0.269 \pm 0.194$ & $ 0.569 \pm 0.138$ & $ 0.721 \pm 0.202$ \\ 
J1330+6519 & 3.27 & 18.69 & 2.951 & 20.80 & 0.14 & 0.54 & $ 0.087 \pm 0.086$ & $ 0.410 \pm 0.146$ & $-0.120 \pm 0.099$ & $-0.228 \pm 0.158$ \\ 
J1354+0158 & 3.29 & 19.20 & 2.562 & 20.80 & 0.17 & 0.42 & $ 0.124 \pm 0.126$ & $ 0.461 \pm 0.218$ & $ 0.050 \pm 0.141$ & $ 0.007 \pm 0.231$ \\ 

\hline
\end{tabular}
\end{center}
\vskip 0.2cm
\noindent
$^a$ Width of the bin in emission redshift adopted for selecting control quasars.\\
\noindent
$^b$ Width of the bin in $z$ magnitude adopted for selecting control quasars.\\
\noindent
$^c$ Mean color and dispersion of the control quasars selected at the same redshift and
magnitude of the DLA-quasar.\\
\noindent
$^d$ Deviation from the mean color of the DLA-quasar.\\
\label{default}
\end{table*}

To construct the  control sample  we started again
from the   list of 5164 DR5 quasars
with $2.25 \leq z_e \leq 3.5$ and 
SNR $\geq$ 4 in the  spectral window 1440-1490\AA. 
We then rejected quasars with spectral signatures of potential sources  of reddening,
using the same automated algorithms adopted for the absorption sample.
In this way we discarded 270 BAL quasars (BAL flag=1 or 2), 
997 quasars with strong \ion{Mg}{ii} lines
(rest frame equivalent width $\geq 1$ \AA), and 669 quasars with strong \ion{H}{i} systems
(including DLA systems closer than 3000 km s$^{-1}$ to the    Ly $\alpha$ emission).

The remaining 3228 quasars   show, in many cases, absorption systems    
undetected by the automated algorithms.
This is due to the fact that the algorithms do not search for  absorption lines
  in spectral regions   with insufficient SNR (e.g.
portions of the Lyman\,$\alpha$ forest particularly crowded)
 or   close to strong quasar emissions (e.g. the \ion{O}{vi} emission). 
While this limitation is not critical for the selection of the absorption sample
(only {\em bona fide} Damped systems  are 
automatically identified), 
it is a reason of concern for the selection of      control quasars
free of any potential source of reddening.
We therefore proceeded to a visual classification of the   quasar  spectra
in a search for intervening absorbers 
not detected  automatically.
Metal lines and strong \ion{H}{i} systems
were tentatively identified by visual inspection
with the following criteria.

Narrow absorption lines redward of the QSO Ly$\alpha$ emission
with central depth $\gsim 0.2$ relative to the adjacent continuum
and located away from the CIV/SiIV quasar profiles
were tentatively attributed to low-ionization metal absorptions.
The SNR preselection was in general sufficient to identify
features with central depth $\gsim 0.2$ in the full spectral range.
For the typical FWHM that can be appreciated by eye in the SDSS spectra ($\gsim$20 \AA)
a central depth of 0.2 corresponds to an equivalent width of $\approx 4$ \AA\ in the observer
rest frame. This corresponds to a rest-frame equivalent width of $\approx 1$ \AA\ 
($\approx 2$ \AA) 
for absorbers at redshift $z_a=3$ ($z_a=1$). Metal lines with equivalent widths of this magnitude are generally saturated and suggestive of strong metal absorptions potentially associated with DLA systems.

Saturated absorptions   in the Ly  $\alpha$ forest
with FWHM $\gsim$ 40 \AA\  and zero residual intensity
were   attributed to strong HI systems. For the typical absorption redshift
detectable in our spectra,  this threshold roughly
corresponds to an equivalent width of $\approx$ 10 \AA\  in the rest frame of the QSO.
This value is traditionally used to identify   candidate Damped Ly$\alpha$ absorptions 
in low resolution spectra
(Wolfe et al. 1986). For features lying on the wings of strong quasar emissions
the criterion was relaxed to FWHM $\gsim$ 20 \AA,
given the fact that part of the absorption is probably washed out by the emission itself.
 
In the course of the visual inspection process we also  searched for  
quasars with strong Ly\,$\alpha$ self-absorption
and  BAL quasars not identified automatically.

As a result of the visual inspection process we rejected 
37 quasars with BAL features,
1074 suspect metal systems\footnote
{
This number is consistent with estimates of the number of  
intervening DLA systems at $z_a \leq 2.2$
not detectable in Ly\,$\alpha$, but potentially detectable as metal systems.
}, 
141 potential DLA systems (including absorptions in the proximity of the Ly\,$\alpha$ emission),
and 17 cases with a strong Ly\,$\alpha$ self-absorption.
In this way
we were left with a sample of 1959 control quasars
listed in Table 2. This is a reasonable approximation of an ideal sample of quasars 
without intervening sources of reddening.
As we explain in the next section, this sample is used as a ``pool''  
from which we extract subsets of control quasars specifically designed
for each quasar of the absorption sample.

\begin{table}[htdp]
\caption{ 
Pool of control quasars used in the present investigation (see Section 2).
The first 20 quasars are shown. The full table is available in electronic form
(see also http://adlibitum.oats.inaf.it/vladilo/tables.html).
}
\begin{center}
\begin{tabular}{cccr}
\hline
SDSS & $z_e$ & $m_r$ & SNR$^a$ \\
 & & [mag] & \\
\hline
073512.52+292017.0 & 2.26 & 19.44 &     6 \\ 
105703.34+051141.5 & 2.27 & 19.44 &     5 \\ 
152715.82+491833.4 & 2.27 & 18.19 &    13 \\ 
104129.27+563023.5 & 2.27 & 18.98 &     9 \\ 
141123.51+004252.9 & 2.27 & 18.06 &    15 \\ 
091209.59-000954.9 & 2.27 & 19.13 &     9 \\ 
131902.33+432235.1 & 2.27 & 18.70 &    11 \\ 
085424.97+543938.7 & 2.27 & 18.84 &    10 \\ 
150158.36+365348.5 & 2.27 & 18.96 &     8 \\ 
160216.74+293038.8 & 2.27 & 18.69 &    11 \\ 
005202.39+010129.3 & 2.27 & 17.40 &    53 \\ 
105852.64+000917.8 & 2.27 & 18.79 &    10 \\ 
140108.59+381417.0 & 2.27 & 19.09 &    12 \\ 
151932.72+481349.9 & 2.27 & 18.87 &     9 \\ 
100636.44+050031.0 & 2.27 & 19.20 &     8 \\ 
153013.94+580802.4 & 2.28 & 18.05 &    18 \\ 
094702.76+321818.4 & 2.27 & 19.73 &     6 \\ 
150806.89+005610.3 & 2.28 & 18.01 &    16 \\ 
123435.50+540323.9 & 2.28 & 18.33 &    14 \\ 
110914.13-003600.7 & 2.28 & 18.96 &    13 \\ 

\hline
\end{tabular}
\end{center}
\vskip 0.2cm
\noindent
$^a$ Mean signal-to-noise ratio in the rest-frame spectral window 1440-1490\AA. \\
\label{default}
\end{table}

  \begin{figure}
   \centering
 \includegraphics[width=8.5cm,angle=0]{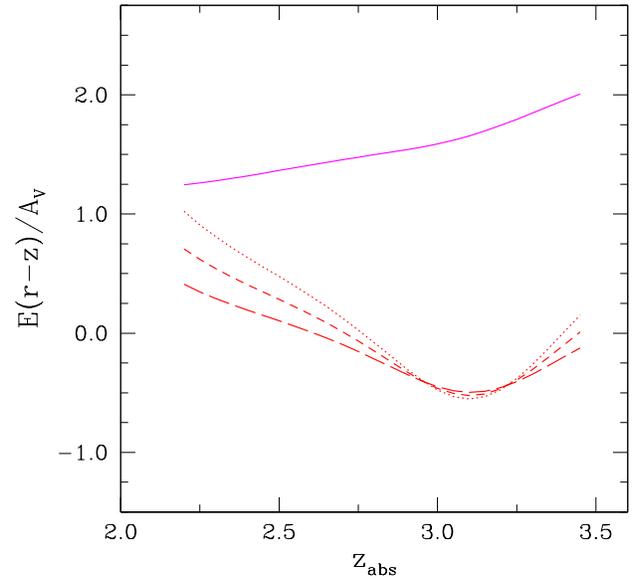}  
  \caption{ 
  color excess $E(r-z)$  in the observer frame 
   calculated  for  an absorber at redshift $z_a$
  with rest-frame extinction $A_V=1$ mag. 
    The lines correspond to  an SMC-type extinction curve 
    (magenta, solid line; Pei 1992, Gordon et al. 2003)  and three MW-type extinction curves (red lines; Cardelli et al. 1998) computed at  
     three values of $R \equiv A_V/E(B-V)$
   (dots: $R=2.5$; short dashes: $R=3.1$; long dashes: $R=4.0$).
  }
              \label{fig-NErz-za}%
    \end{figure}

\section{The method}

\subsection{Derivation of the mean color excess}

Given two measurements $m_x$ and $m_y$ of the apparent magnitude of a quasar
in two photometric bands $x$ and $y$,
we call $C(y-x)=m_y-m_x$ the {\em quasar color}. 
If dust is present in a foreground absorption system,
the observed color will be different from the  intrinsic,
unreddened color $C_0(y-x)$, yielding a
{\em color excess}
$E(y-x)=C(y-x)-C_0(y-x)$. This
cannot be measured directly since 
the unreddened color of the individual quasar is unknown.
We can however measure the mean color excess of a sample
of DLA-quasars in the following way.
For each DLA-quasar we build a set of $n_j$ unreddened control quasars 
with same redshift and magnitude.
We  then estimate the mean color of these unreddened quasars,
$\langle C^u(y-x)  \rangle$, 
and hence the
{\em deviation from the mean color}
 \begin{equation}
\Delta(y-x)
 = C(y-x) - \,  \langle C^u(y-x)  \rangle    \,  .
\label{DELTA1}
\end{equation}
For a large sample of $n_i$ DLA-quasars we measure the {\em mean color excess}  
 with the expression
\begin{equation}
 \langle  E (y-x) \rangle = \langle \Delta (y-x) \rangle    ~,
 \label{equality}
\end{equation}
where $ \langle  \Delta (y-x) \rangle$ %
is the average of the $n_i$ values $\Delta_i$.
The validity of Eq. (\ref{equality}) is discussed in Appendix A.

  \begin{figure}
   \centering
 \includegraphics[width=8.5cm,angle=0]{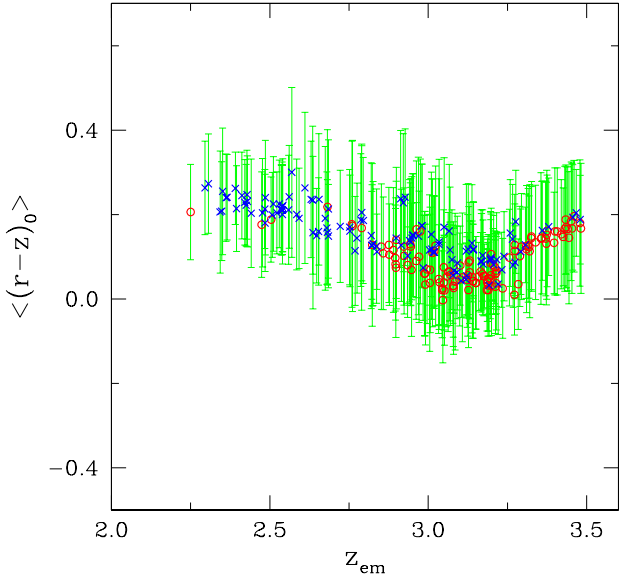}  
  \caption{ 
  Mean color ${\langle (r-z)_0 \rangle}$ of the QSOs  control samples
  used to estimate the unreddened colors of the DLA-QSOs.  
  The mean color is plotted versus the QSO redshift, $z_e$.
  Bright ($r < 19.1$ mag) and faint ($r \geq 19.1$ mag) QSOs are plotted with
  blue crosses and red circles, respectively.
  }
              \label{fig-Qsocolors}%
    \end{figure}

\subsection{Choice of the color index}

Since our ideal goal is to measure the color of the quasar continuum,
we excluded photometric bands 
contaminated by the absorptions in the Ly $\alpha$ forest
or the   Ly $\alpha$ emission of the quasar. 
Taking into account the average wavelengths $\overline{\lambda}$ and the FWHMs of the 
SDSS photometric bands (Fukugita et al. 1996; 
 see also SDSS web page
http://www.sdss.org/dr5/instruments/imager) 
this condition constrains 
the maximum quasar redshift to $z_{e,\mathrm{max}} = 
[ ( \overline{\lambda}  - { 1 \over 2} \, \mathrm{FWHM})  / 1216 ] - 1
 = 1.67, 2.29, 3.50, 4.52,$ and 5.78 
for the $u$, $g$, $r$, $i$, and $z$ bands, respectively. 
Since the wavelength coverage of SDSS spectra prevents detection of DLA systems
at $z_a \lsim 2.2$, 
only the $r$, $i$ and $z$ bands provide quasars with 
uncontaminated photometry (in the sense specified above)
and sufficient redshift coverage for detection of DLA systems. 
The maximum leverage for reddening detection with these bands is given by the $(r-z)$ color.

\subsubsection{The $(r-z)$ color index}

In Fig. \ref{fig-NErz-za} we plot the color excess $E(r-z)$ in the observer's frame
that we expect to 
measure for an absorber at redshift $z_a$ with rest-frame extinction $A_V=1$ magnitude.
To estimate $E(r-z)$  we considered different possible
types of extinction curves for the dust in the DLA systems, such as
the Milky-Way curves  by Cardelli et al. (1998), with the characteristic extinction bump at 2175\AA,
and the SMC curves (Pei 1992, Gordon et al. 2003), characterized by a more regular UV rise. 
The extinction curves are
normalized in the $V$ band,  
$\xi(\lambda) \equiv A_\lambda/A_V$,
all the quantities in this definition
being in the rest frame of the absorber.
For an absorber at redshift $z_a$ the extinction measured at wavelength $\lambda^\mathrm{obs}_x$ in the observer's frame  is $A_{\lambda^\mathrm{obs}_x} = A_V \, \xi({\lambda^\mathrm{obs}_x \over 1+z_a})$. 
The color excess in the observer's frame per unit extinction in the rest frame 
is therefore
\begin{equation}
{E(\lambda^\mathrm{obs}_y-\lambda^\mathrm{obs}_x) \over A_V}
=
\delta \xi \, (z_a ; \lambda^\mathrm{obs}_x, \lambda^\mathrm{obs}_y) 
\label{Erz_efficiency}
\end{equation}
where
$
\delta  \xi \, (z_a ; \lambda^\mathrm{obs}_x, \lambda^\mathrm{obs}_y) =
\xi({\lambda^\mathrm{obs}_y \over 1+z_a})-\xi({\lambda^\mathrm{obs}_x \over 1+z_a}) ~.
$
From this expression we have computed the curves shown in Fig. \ref{fig-NErz-za}.

For an SMC-type extinction curve we expect
a smooth variation of the predicted color excess in the redshift interval of interest,
 with a typical value  $E(r-z) \simeq 0.15$ mag
 for absorbers with rest-frame $A_V = 0.1$ mag.
For a MW-type curve we expect a small (or even negative)  reddening
in the interval of absorption redshift of our sample.
The minimum in the   MW lines of Fig.  \ref{fig-NErz-za} 
corresponds
to minimum between the end of the extinction bump at 2175\AA\
and the UV rise. 
The implications of these different behaviours
 of the SMC-type and MW-type extinction curves are discussed in Section 5.1.

\subsubsection{The $(g-z)$ color index}

The measurement of the $(g-z)$ color is challenging
because the $g$ band is generally contaminated by the 
Ly $\alpha$  forest, or even the QSO Lyman break,
in  the interval of emission redshift considered.
However, measuring the reddening in the $(g-z)$ color,
in addition of the $(r-z)$ color, offers several advantages. 
One is the higher leverage for reddening detection:   we expect a gain of
a factor of $\simeq 2.2$ in the ratio $E(g-z)/E(r-z)$
at the redshift of our sample if the extinction curve is of SMC type.
 Another advantage is the possibility of probing the
wavelength dependence of the reddening, i.e. the extinction curve of the dust,
by comparing the $(g-z)$ and $(r-z)$ color excesses.
We decided therefore to perform the measurements also in the $(g-z)$ color index
taking care of the contaminations present in the $g$ bandpass.

To avoid  overlap of the $g$ band with the QSO Lyman break, we restricted the analysis  
of the $(g-z)$ colors to the QSOs with $z_e \leq 3.4$  in our sample.
We then took care of the contamination due to Ly $\alpha$ absorptions.
The contamination due to the forest is present both in the
quasars of the absorption sample and of the control sample.
Line-to-line variations of the forest absorption will
contribute to the dispersion of the quasar colors, making
the measurement of the color excess more difficult.
If such variations are randomly distributed,
we still can measure the difference between the quasar colors  
of the absorption and control samples. 
In doing this, however, we must take into
account the fact that the $g$ band of the absorption sample
is generally contaminated by the damped Ly\,$\alpha$ profile, which is absent in the control quasars.
To cope with this problem we correct the $g$ magnitude by taking into account
the effect of the damped Ly\,$\alpha$ absorption\footnote
{ A similar procedure was adopted by Ellison et al. (2005) to correct
for the flux suppression due to Ly\,$\alpha$ absorption in the $B$ band. 
Here we consider the flux suppression due to all the lines of the Lyman series
that overlap the SDSS $g$ band.}
in the filter bandbass.
The magnitudes in the SDSS photometric system are defined as
$m = -2.5 \, \log \int d (\log \, \nu) \, f_\nu \, S_\nu + \mathrm{constant}$ (Fukugita et al. 1996),
where $f_\nu$ is the quasar spectral distribution and $S_\nu$  the response function
of the filter.
We define the correction for the DLA absorption in the $g$ band as
\begin{equation}
\delta m_g = -2.5 \, \log 
{
\int d (\log \, \nu) \, f^\circ_\nu \, S^{(g)}_\nu
\over
\int d (\log \, \nu) \, f^D_\nu \, S^{(g)}_\nu
}
~,
\label{gCorr}
\end{equation}
where 
$f^\circ_\nu$ is the quasar spectral distribution
without DLA absorption,
$f^D_\nu$ the spectral distribution with   
DLA absorption,  and
$S^{(g)}_\nu$ the response function of the  $g$ filter (Fukugita, 2006, priv. comm.).
To estimate $\delta m_g$ we model
$f^\circ_\nu$ with a power law
with spectral index $\alpha_\nu \simeq -0.5$, based on measurements
of SDSS spectra of quasars in the redshift interval of interest for the present work
(Vanden Berk et al. 2001, Desjacques et al. 2007).
For each DLA system of our sample 
we then compute the theoretical absorption profile of the
lines of the Lyman series, $\ell_\nu$, 
with intensity and position of the lines specified by the 
  \ion{H}{i} column density and redshift.
From this we compute $f^D_\nu = \ell_\nu \times f^\circ_\nu$
and hence the correction $\delta m_g$, which is finally added to the observed magnitude.
The theoretical Voigt profiles are computed with the 
 FITLYMAN routines (Fontana \& Ballester 1995)
included in the ESO MIDAS software package.
The typical value of $\delta m_g$ is of $\simeq -0.05$ mag, with values
as low as $-0.006$ mag and as high as $-0.17$ mag in the most extreme cases.

\subsection{Implementation of the method}

The ``BEST'' $ugriz$ imaging data were recovered
for all quasars of the absorption and control samples. 
These photometric data were  
corrected  for the effects of the Galactic extinction according to the
prescriptions given by Schneider et al. (2005).

For each quasar of the absorption sample we  
calculate the colors $(r-z)_i$ and $(g-z)_i$ and then
select a taylored subset of control
 quasars closest in redshift and magnitude 
in the following way.
We first select  the $n_{z_e}$ control quasars closest to the redshift $z_e$
of the DLA quasar.
From this subset we then select the $n_m$ control quasars
closest to the $z$-band magnitude $m_z$ (the least affected by extinction)
of the DLA quasar. 
This last subset is used to estimate the mean 
unreddened colors
${\langle (r-z)_0 \rangle}_i$ and ${\langle (g-z)_0 \rangle}_i$.
The deviations from the mean color  are finally calculated as
$\Delta_i(r-z) = (r-z)_i - {\langle (r-z)_0 \rangle}_i$ and
$\Delta_i(g-z) = (g-z)_i - {\langle (g-z)_0 \rangle}_i$.

With this procedure each subset of control quasars has
the same size $n_m$ for all DLA-QSOs and we have
a comparable statistics for all the measurements.
The choice of $n_{z_e}$ and $n_m$ is determined by the requirement
that the total intervals in emission redshift, $\delta_i(z_e)$, 
and $z$ magnitude, $\delta_i(m_z)$, spanned by each subset are
sufficiently small to guarantee a good degree of homogeneity of 
the control samples.
In practice, this gives the contraints $n_{z_e} \lsim 100$ 
and $n_{z_e}/n_m \gsim 4$ in order to
have typical (median) values $\delta_i(z_e) \simeq 0.1$  
and  $\delta_i(m_z) \simeq 0.5$ mag.
The redshift interval is similar to the redshift bins
commonly adopted in studies of the mean quasar color 
as a function of $z_e$ (e.g. Richards et al. 2001). 
 Changes of the intrinsic slopes of the quasar continua over
the above magnitude interval are expected to be   modest
(Yip et al. 2004). 

In Table 1 we report the measurements of    
$\Delta_i(r-z)$ and $\Delta_i(g-z)$
for the DLA-QSOs of the absorption sample.
In  columns 6 and 7 of the table we give
the intervals in emission redshift, $\delta_i(z_e)$, and $z$ magnitude, $\delta_i(m_z)$, 
spanned by each subset of control quasars obtained by adopting $n_{z_e}=100$ and $n_m=25$.
In columns 8 and 9 we list the mean colors 
 $\langle (r-z)_0 \rangle_i$ and $\langle (g-z)_0 \rangle_i$, obtained from 
the weighted mean of the colors of each control subset.
The weights are taken to be proportional to the inverse squares of the errors
of individual colors.

The errors of individual colors of DLA quasars and control quasars 
are obtained by propagation of the
photometric errors and of the errors of the Galactic extinction correction.
For the latter we adopted an uncertainty of $\pm 50\%$ of the correction itself.
The $(g-z)$ data include the correction for damped Ly  $\alpha$ absorption in
the $g$ band, when approprite. An uncertainty of $\pm 50\%$ of this correction is propagated
in the error budget.

The  errors of  the mean colors ${\langle (r-z)_0 \rangle}_i$  
and $\langle (g-z)_0 \rangle_i$ quoted in Table 1
are the standard deviation of the colors of each subset.
These errors 
dominate the budget of the errors of the color deviations $\Delta_i(r-z)$
and $\Delta_i(g-z)$.

In Fig. \ref{fig-Qsocolors} we plot the mean colors ${\langle (r-z)_0 \rangle}_i$
of the subsets of unreddened quasars versus $z_e$.
Each data point corresponds to the subset of an individual DLA-quasar of Table 1.
We use different symbols for  quasars brighter and fainter than the median
magnitude of the absorption sample, $m_r = 19.1$.
In addition to a trend with $z_e$, known from previous studies, the figure shows
 a dependence on the quasar magnitude.
For instance, at $z_e \approx 3$, where many quasars
both fainter and brighter than $m_r = 19.1$  are present,
the fainter quasars   (red circles) lie systematically below the brighter
ones (blue crosses).  
This result justifies our choice of grouping the   control quasars  
  not only in redshift, but also in magnitude.

\begin{table*}[htdp]
\caption{Mean reddening in the observer's frame, 
$\langle E \rangle$, and mean extinction in the rest frame,
$\langle A_V \rangle$,  for different colors and DLA/QSO samples.}
\begin{center}
\begin{tabular}{| c|cccc|cccc | }
\hline 
color & & & $(r-z)$   & & & & $(g-z)$     & \\
\hline 
Redshift interval & & & $2.25 \leq z_e \leq 3.5$   & & & & $2.25 \leq z_e \leq 3.4$     & \\
\hline
Absorption
 &   $n$   & ${\langle E(r-z) \rangle}$ & $\langle A_V \rangle$ & $\langle A_V/N(\ion{H}{i}) \rangle$  
 &   $n$   & ${\langle E(g-z) \rangle}$ & $\langle A_V \rangle$ & $\langle A_V/N(\ion{H}{i}) \rangle$  \\
   sample           &           & [10$^{-3}$ mag]& [10$^{-3}$ mag]   & [10$^{-23}$ mag cm$^2$]          
   &           & [10$^{-3}$ mag] & [10$^{-3}$ mag]  & [10$^{-23}$ mag cm$^2$] \\
\hline
1$^a$  &  248   & 27 $\pm$ 8.8 & 19 $\pm$ 5.9 & 4.1 & 232 & 54 $\pm$ 12  &  17 $\pm$ 3.8 & 3.2 \\ 
2$^b$  & 247 & 24 $\pm$ 8.3 & 17 $\pm$ 5.6 & 3.5 & 231 & 51 $\pm$ 12 & 16 $\pm$ 3.7 & 2.8\\
3$^c$  & 246 & 22 $\pm$ 8.0 & 16 $\pm$ 5.4 & 3.0 & 229 & 48 $\pm$ 11 & 15 $\pm$ 3.5 & 2.4\\ 
\hline
Bootstrap$^d$ & 248 & 27 $\pm$ 9.2 &  &  &  232 &  54 $\pm$ 14 & & \\
\hline 
\end{tabular}
\end{center}
\vskip 0.2cm
\noindent
$^a$ Complete absorption sample of Table 1. \\
\noindent
$^b$ Cases with $\Delta_i  > \langle \Delta \rangle + 5 \times \sigma$ are excluded from the
complete sample
($\sigma = \sigma_m \times \sqrt{n}$). \\
\noindent
$^c$ Cases with $\Delta_i  > \langle \Delta \rangle + 3 \times \sigma$ are excluded from the
complete sample.\\ 
\noindent
$^d$ Mean and standard deviation of the mean color excess of 10,000 bootstrap samples obtained from the
absorption sample 1. \\

\label{default}
\end{table*}%

\section{The mean color excess}  

In Table 3 we give
the weighted  means ${\langle E(r-z) \rangle}$
and ${\langle E(g-z) \rangle}$ 
estimated with the expression ${\langle E \rangle} = \langle \Delta \rangle =
\sum_i w_i \Delta_i$,
where  
$w_i=\sigma_i^{-2} / \sum_{i} \sigma_i^{-2}$, and  $\sigma_i$ are
the errors of individual measurements. 
%
The use of the weighted mean allows us to minimize the
contribution of the  QSOs with largest uncertainty
of their intrinsic colors, the dominant source of the error budget.

For the error  of the mean we adopted the unbiased estimate  
 \begin{equation}
 \sigma_m = \sqrt{  \sum_i w_i \, 
 (\, \Delta_i -  {\langle \Delta \rangle} \, )^2   \over  (n-1)  }
 \label{WME}
 \end{equation}
(Linnik 1961) with the normalized weights $w_i$ given above.
This expression is valid 
if the measurements of $\Delta_i$ are uncorrelated.
We performed two tests to make sure that correlated errors  
 are unimportant and therefore the adopted statistical error $\sigma_m$
is reliable.
%
First we computed the covariance matrix
and compared the strength of off-diagonal terms with that of diagonal terms.
The covariance matrices of the $\Delta_i(r-z)$ and $\Delta_i(g-z)$ data samples
were computed using a bootstrap method over
60 random realizations of each sample.
We found that the normalized covariance matrices are fairly close to be diagonal,
with off-diagonal terms tipically around a few percent, and always below 25\%.
The low values of the off-diagonal terms indicate that there is little correlation
between the measurements. 
Then we studied the behaviour of  $\sigma_m$ as a function of the sample size.
This can be done with a  random grouping technique described
by Hampel et al. (1986; Chap.8). 
From this test we find that the adopted error
follows the theoretical $n^{-0.5}$ decrease,
 as expected for uncorrelated measurements.  
The adopted error of the mean is therefore a good estimate
of the statistical error of $\langle E \rangle$.

\subsection{Confidence level of detection}

From   the results shown  in Table 3  one can see that
the mean $(r-z)$ reddening is detected at    
$\simeq 3 \, \sigma_m$ confidence level (statistical error)
in the complete sample of 248 DLA/QSO pairs 
for which the $r$  band  is uncontaminated by Ly  $\alpha$ forest
($z_e \leq 3.5$).
The mean $(g-z)$ reddening is detected at $\simeq 4.5 \, \sigma_m$ confidence level  
in the sample of 232 DLA/QSO pairs 
with $g$ magnitude  uncontaminated by the QSO Lyman break  ($z_e \leq 3.4$)
and  corrected for DLA flux suppression   
(Section 3.2.2).

\begin{figure*}
   \centering
 \includegraphics[width=8.5cm,angle=0]{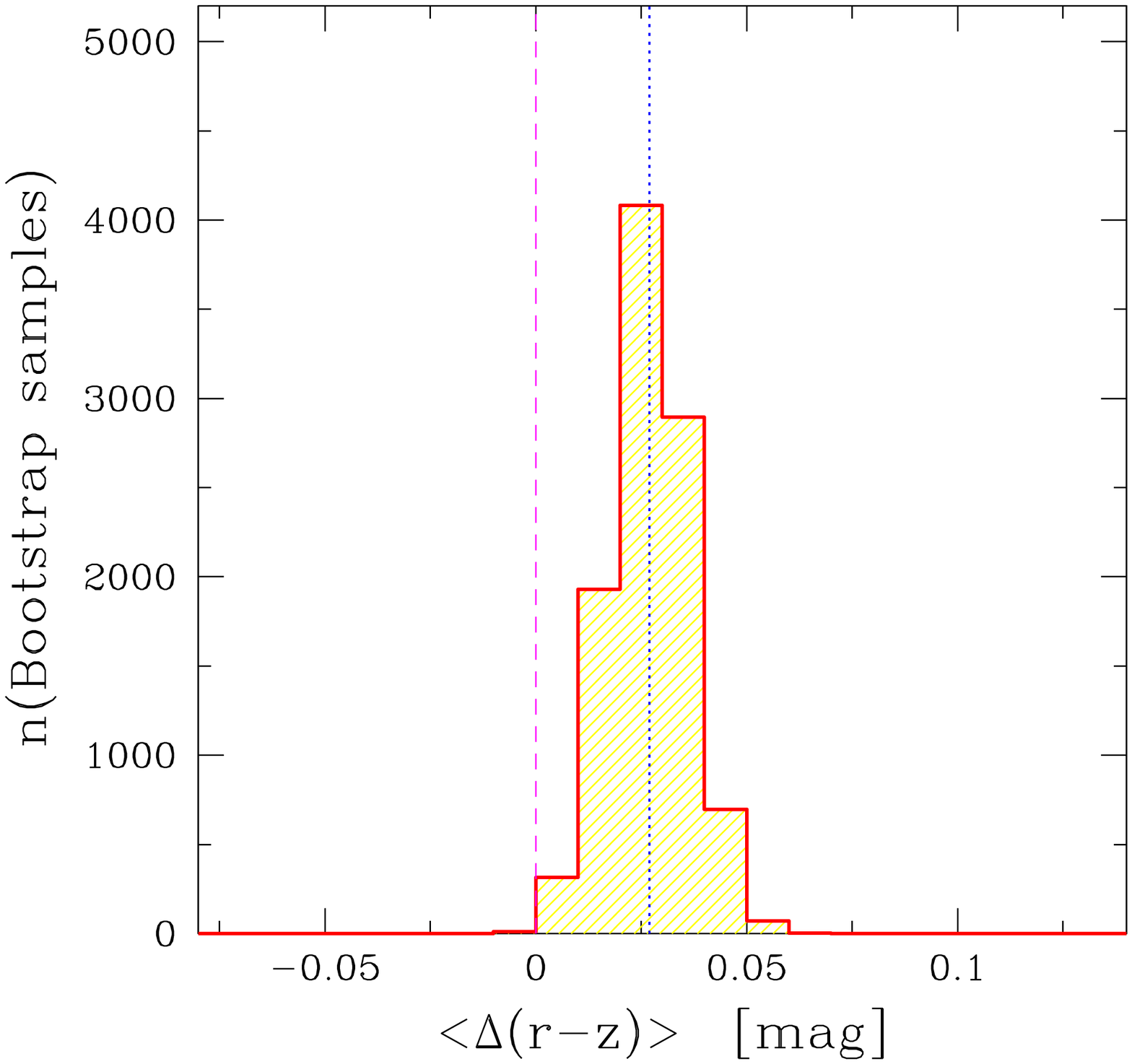}  
 \hskip 0.5cm
 \includegraphics[width=8.5cm,angle=0]{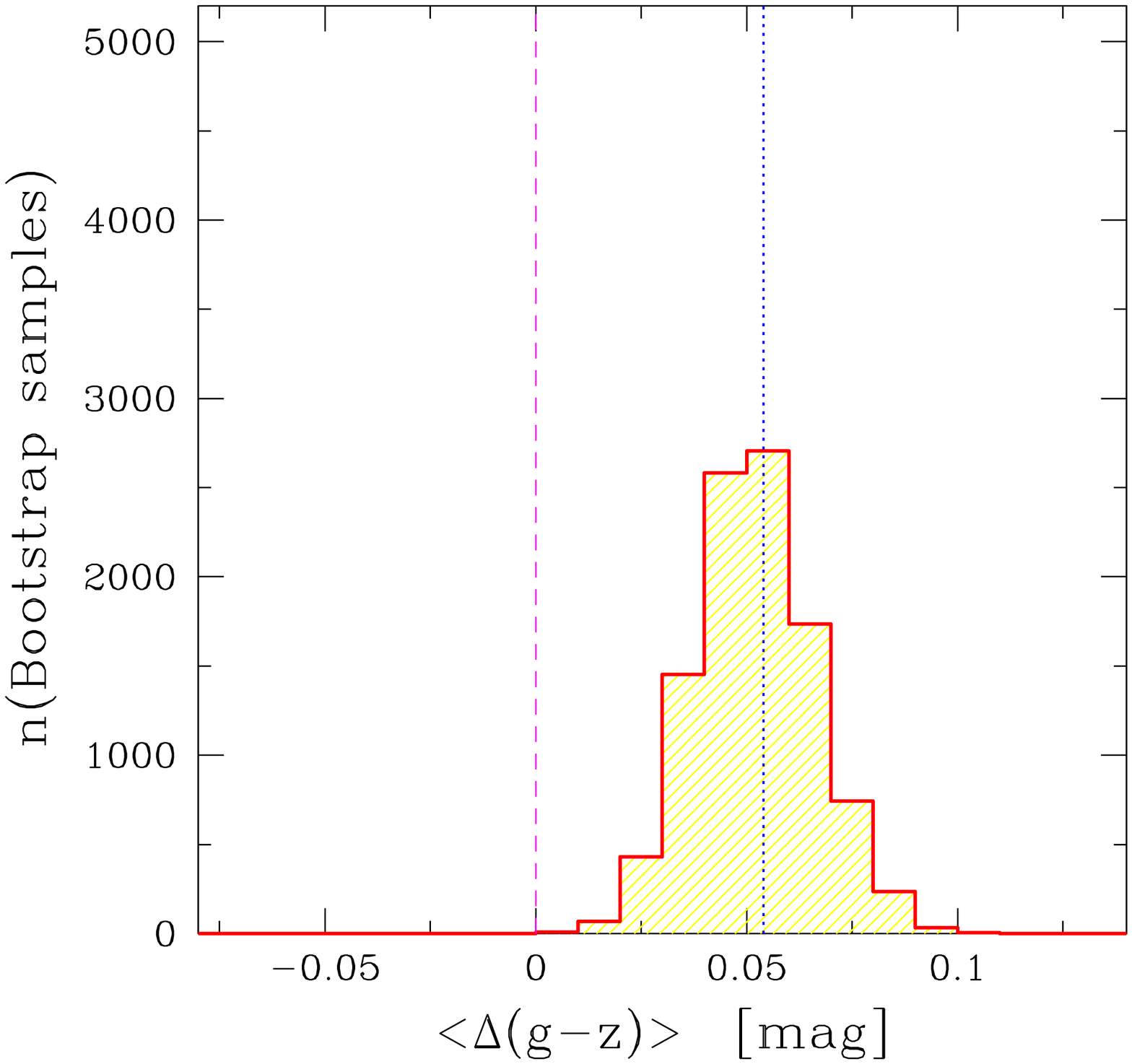}  
  \caption{ Frequency distribution of the mean color excess 
  $\langle \Delta(r-z) \rangle$ and $\langle \Delta(g-z) \rangle$ measured in 10,000
   bootstrap samples obtained from the list of $\Delta$ values of 
   Table 1. Dotted line: mean color excess of the original sample. See  Section 4.1. 
  }
              \label{bootstrap}%
 \end{figure*}

To estimate the confidence level of the detection other
than in terms of $\sigma_m$  
we applied the method of bootstrap resampling (Efron 1979).
%
A bootstrap sample is obtained by extracting $n$ data at random {\em with repetition}
from the original sample of $n$ measurements. This type of extraction is, in practice,
an extraction {\em with replacement} because
part of the original measurements is replaced by repeated data.
By iterating this extraction process many times
it is possible to build a  
large number $N_B$ of bootstrap samples 
from the original set of $n$ data.
For each of the $N_B$ samples one can compute a statistical estimator of interest
(e.g. the weighted mean) and study the distribution of such estimator, 
without making any  assumption  on the parent distribution.
%
%
We applied this method to  
the original sample of $n$ measurements   $\Delta_i$, from which
  we built up  $N_B = 10,000$  bootstrap samples. 
The weighted  mean $\langle \Delta \rangle$ was recomputed in each case. 
We analyzed the resulting distribution of $\langle \Delta \rangle$ values
(Fig. \ref{bootstrap})
to estimate the fraction of cases in which the mean value is positive. 
We find    $\langle \Delta(r-z) \rangle > 0$
in $\simeq 99.90\%$ of the $\Delta_i(r-z)$ bootstrap samples
and $\langle \Delta(g-z) \rangle > 0$ in $\simeq 99.99\%$ of the $\Delta_i(g-z)$
bootstrap samples.
These percent figures indicate the probability that a positive
color excess has been detected.

\subsection{Analysis of ``twin control samples''}

\begin{figure*}
   \centering
 \includegraphics[width=8.5cm,angle=0]{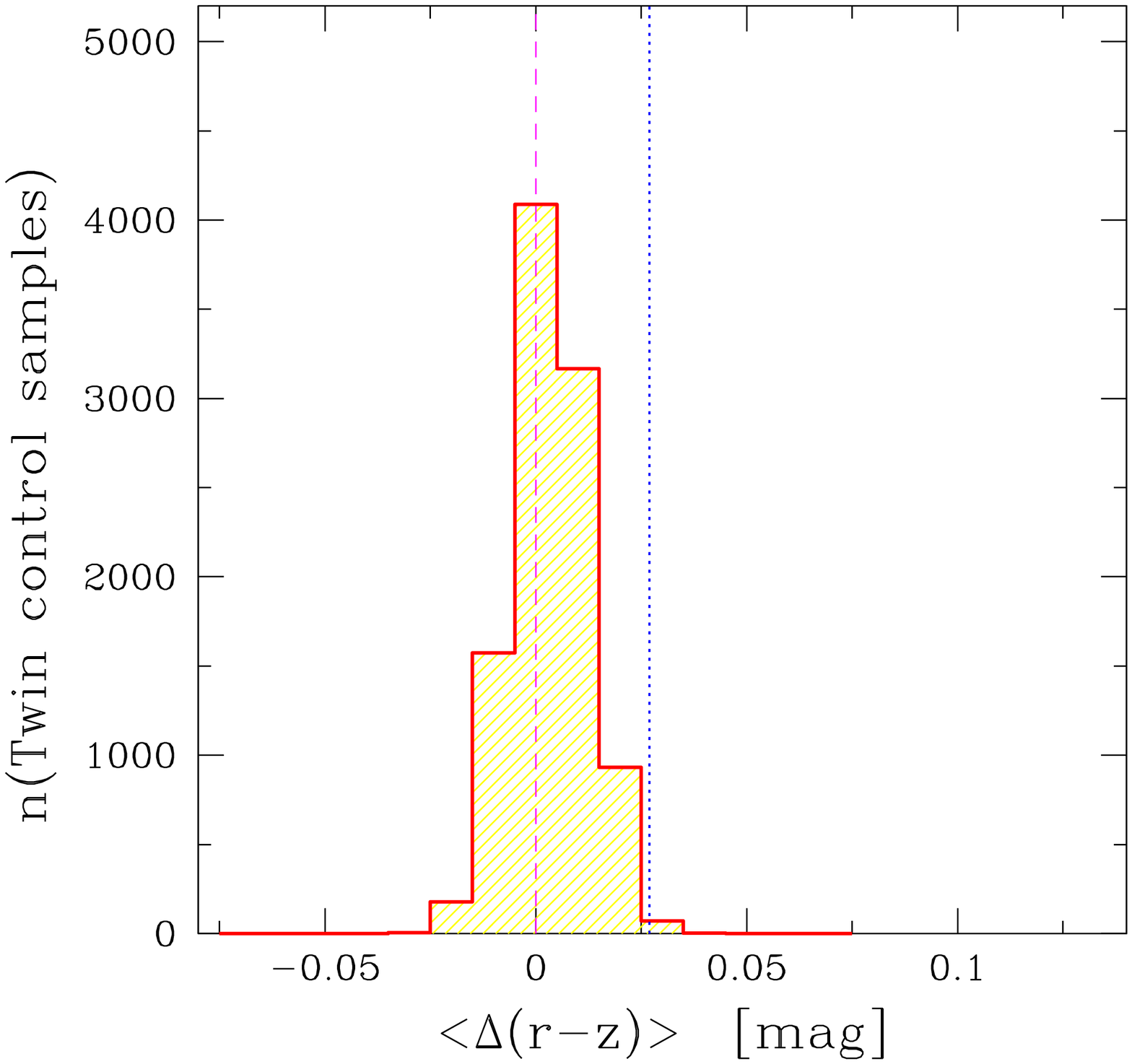}  
 \hskip 0.5cm
 \includegraphics[width=8.5cm,angle=0]{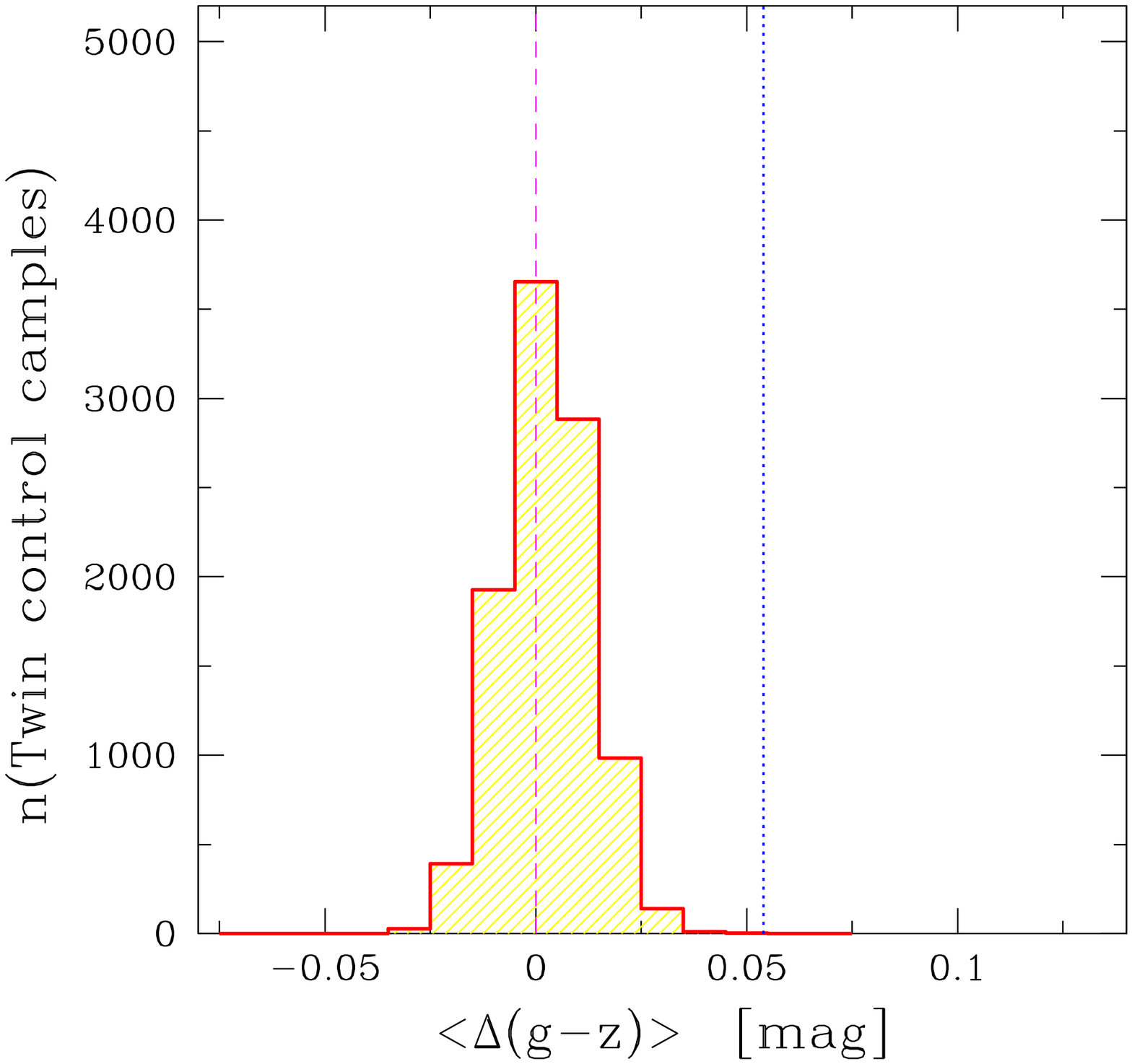}  
  \caption{ Frequency distribution of the mean deviations 
  $\langle \Delta(r-z) \rangle$ and $\langle \Delta(g-z) \rangle$ measured in 10,000 
   control samples
  with identical size, redshift distribution and magnitude 
  distribution of the absorption sample.
  Dotted line: mean color excess of the absorption sample. See  Section 4.1. 
  }
              \label{fig-Twins-distr}%
 \end{figure*}

A reason of concern in the present analysis is the possibility that
the mean color excess that we detect is due to sources of reddening other than the DLA systems
(e.g. dust in the quasar environment or in unidentified absorbers). 
If present, these reddening contribution would also affect the quasars of the control sample.
To understand if  these effects may accidentally yield 
a mean color excess  comparable to that measured in the DLA sample,
we applied our procedure to 
a large number of ``twin control samples''
with same size, redshift distribution and magnitude distribution  
as the DLA sample.
We then analyzed the values of $\langle \Delta \rangle$
obtained from these  twin control samples  in a search for
cases with a value equal to those reported in Table 3.
In practice,
each  twin control sample was created by extracting at random,
for each DLA-QSO of  Table 1, a control QSO 
at the same redshift and magnitude.
%
The control QSOs extracted in this way were treated as   DLA-QSOs and temporarily
excluded from the  estimate of
the  mean unreddened colors $\langle C^u \rangle$  (Section 3.1). 
As a consequence, the  estimates of $\langle C^u \rangle$
were different for each realization of the twin control sample.
We built up 10,000 twin control samples and
computed the weighted mean $\langle \Delta \rangle$ for each of them. 
The resulting frequency distribution  of  $\langle \Delta \rangle$ values
is shown in Fig. \ref{fig-Twins-distr} for both color indices of interest.
For comparison we show the mean color excess of the absorption sample (dotted lines).
The probability for a twin control sample to attain a value 
$\langle \Delta \rangle$
as high as the mean color excess 
 of the absorption sample is $< 0.01\%$ in $(g-z)$    
and $\simeq 0.4\%$ in $(r-z)$.
We conclude that the mean color excess detected in the absorption sample
cannot be ascribed to 
dust in the quasar environment or
in unidentified low-redshift absorbers. 
The same low probability applies to any sort of systematic effect that
might accidentally conspire to yield a mean color excess 
as a consequence of the application of our procedure to
a quasar sample with same size, redshift distribution and magnitude distribution
as the DLA sample.

\subsection{Rejection of the most reddened quasars}

For a correct interpretation of the present results
it is important to understand whether the mean reddening that we detect 
is a signature of the bulk of the DLA population or is just  due
to the presence of a few reddened quasars.
To clarify this point we rejected  the most reddened DLA-QSOs 
from the original sample of Table 1 and repeated the computation of the mean reddening. 
In practice,
we discarded the cases with individual color deviation $\Delta_i$
larger than the mean color deviation $\langle \Delta \rangle$
at  $5 \, \sigma$ level
(``absorption sample 2'') and 
at $3 \, \sigma$ level (``absorption sample 3'').
In the second and third row of Table 3 we show
the mean color excess recomputed 
after rejecting the most reddened cases in this way.
One can see   that
the mean color excess is still detected at $\simeq$ 2.8 $\sigma_m$ level  in $(r-z)$ and 
at  3.5 $\sigma_m$ level in $(g-z)$
for the absorption sample 3. The detection is firmer for the absorption sample 2.

\subsection{Size of the subsets of control quasars}

As an final test we checked the stability of the results
for the adoption of different values of $n_{z_e}$ and $n_m$ 
(Section 3.3). 
We tested this   effect by varying these
parameters  around  their optimal values $n_{z_e} \simeq 100$
and $n_{z_e}/n_m \simeq 4$. 
The resulting changes of the mean reddening are  significantly
smaller than the statistical error discussed above.
  
\section{Discussion}

The analysis presented in the previous section indicates that  
high-redshift quasars with intervening DLA systems
at $2.2 \leq z_a \leq 3.5$ have redder colors than quasars
at the same redshift  
without foreground absorption systems.

The most natural process to explain the origin of this reddening
is dust extinction, with its characteristic rise at shorter wavelengths.
If this is the case, we expect       
the measured reddening to show similarities    
with that predicted for typical interstellar dust extinction curves.
We also expect 
the most reddened quasars to be dimmed by dust extinction and
therefore to be statistically fainter than unreddened quasars.
In Sections 5.1 and 5.2 below we present observational evidence 
consistent with these expectations. 

The location of the dust is of special interest in the present work.
Reddening sources in the quasar environment  cannot explain the difference
in color   between the quasars of the absorption and control samples
since this type of reddening 
would affect in the same way, on the average, the quasars of both samples.
Dust embedded in the DLA systems is
the most natural hypothesis to explain the measured color excess.
This hypothesis bears some predictions that can be tested empirically. 

One is that the reddening should increase with
the column density of the metals embedded in the DLA systems.
Another is that  the dust-to-gas ratio should approximately
scale with the level of metallicity of DLA systems. 
In Sections 5.3 and 5.4 below we present observational evidence
consistent with these expectations. In Sections 5.5 and 5.6 we discuss 
the reddening versus \ion{H}{i} column density and versus absorption redshift.

\subsection{Dust extinction curve}

One of the properties used to characterize
the interstellar dust is its extinction curve,  $\xi(\lambda) \equiv A_\lambda/A_V$.
In spite of strong spatial variability, interstellar extinction curves 
can be classified in two types:
Milky-Way type curves, with the characteristic extinction bump at 2175\,\AA\
(e.g. Cardelli et al. 1998),
and   SMC-type curves, characterized by a  more regular UV rise
(e.g. Pei 1992, Gordon et al. 2003).
The different slopes of these curves can be used to distinguish  between them. 

If the reddening originates in the DLA systems,
we can measure the slope of the DLA dust extinction curve
 by comparing
the mean reddening in the $(g-z)$ and $(r-z)$ colors.
We use the restricted sample of 232 DLA-QSOs with $z_\mathrm{em} < 3.4$,
for which both $(g-z)$ and $(r-z)$ can be derived.
For this sub-sample   we measure
$\langle E(r-z) \rangle = 25 (\pm 9.0)$ 10$^{-3}$ mag 
and we obtain
$\langle E(g-z) \rangle / \langle E(r-z) \rangle =$  2.2  $\pm$ 0.9.
This measurement is in good agreement with the ratio
predicted for absorbers 
with SMC-type extinction curve at $2.2 \leq z_a \leq 3.5$,
$\langle E(g-z)/E(r-z) \rangle^\mathrm{SMC}_\mathrm{pred} \simeq 2.2$.
 
The fact that we detect reddening in the $(r-z)$ color argues
againts a MW-type extinction curve. In fact, 
this type of curve  is expected to give a   negligible 
or even negative reddening
in the interval 
of absorption redshift spanned by of our sample  
(Fig. \ref{fig-NErz-za}).
This is due to the fact that in this interval
 the $(r-z)$ color   probes
a part of the MW extinction curve with negative slope, namely
the end of the 2175\,\AA\ extinction bump.
The fact that we detect reddening in the $(r-z)$ color 
indicates that   extinction curve of MW type must be uncommon in the sample. 


If DLAs with MW-type curves exist, a broad absorption
centered at $(1+z_a) \times 2175\,\AA$
should affect the quasar continuum. We searched for 
 this feature in all the spectra of the absorption sample, and in particular
 of the quasars with anomalous colors, but without success
(see e.g. Fig.  \ref{fig-redquasars}). 

The above results are in line with the general finding that MW-type extinction
curves are rare among quasar absorbers (Junkkarinen et al. 2004, Wang et al. 2004),
while SMC-type are common (Wild \& Hewett 2005, York et al. 2006).
By adopting an SMC extinction curve, our results imply a rest-frame extinction
$\langle A_V \rangle \simeq 0.015/0.019$ mag 
 (Table 3),
corresponding to 
$\langle E(B-V) \rangle \simeq 0.005/0.007$ mag.

      \begin{figure*}
   \centering
 \includegraphics[width=8.5cm,angle=0]{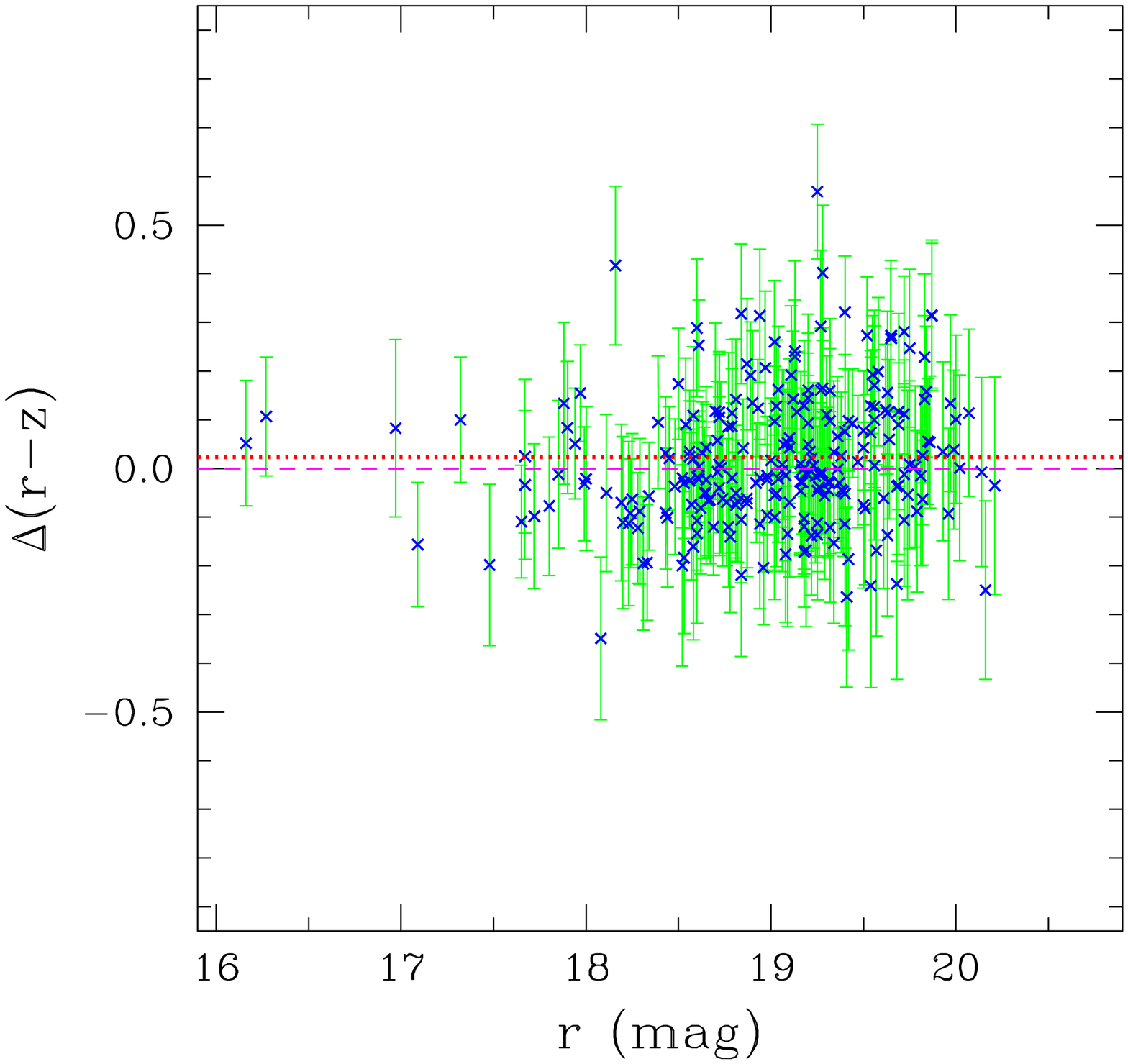}  
 \hskip 0.5cm
 \includegraphics[width=8.5cm,angle=0]{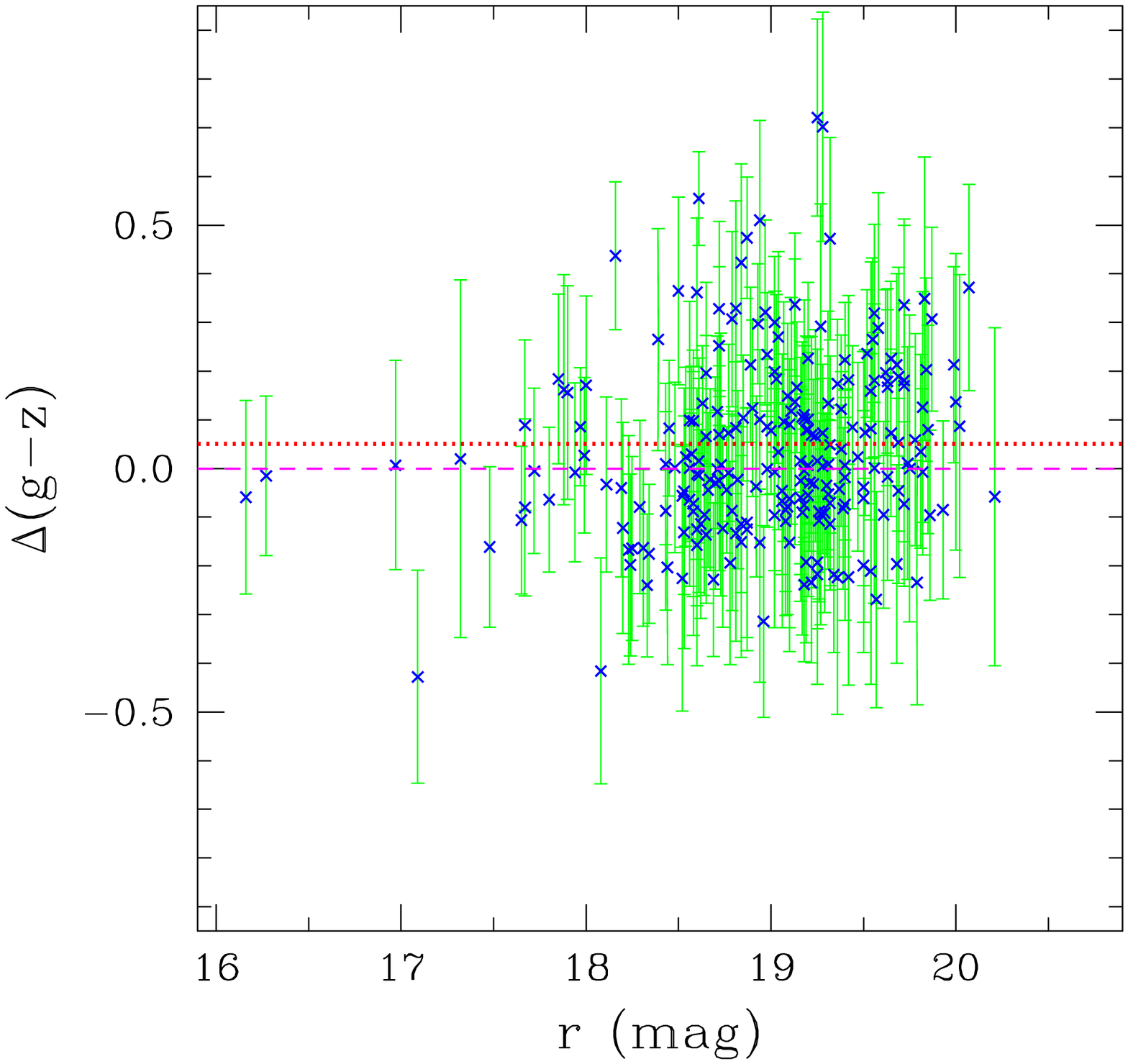}  
  \caption{ Deviation from the mean color, $\Delta(r-z)$ and $\Delta(g-z)$, 
  versus $r$ magnitude for sample 2 of Table 3.
   Dotted lines: mean color excess of the same sample.
  }
              \label{fig-Drz-mr}%
    \end{figure*}

\subsection{Reddening versus quasar magnitude}

As we mentioned above, we expect  
the most reddened quasars to be statistically fainter than unreddened quasars
if the reddening originates in  dust.
In Fig. \ref{fig-Drz-mr} we plot 
$\Delta(r-z)$ and $\Delta(g-z)$ versus the $r$ magnitude. 
The largests values of color excess lie
at faint magnitudes ($r \gsim 18$ mag), 
in broad agreement with this expectation. 
To quantify the effect we compared 
the mean magnitude of the 30 DLA-QSOS with highest  $\Delta(r-z)$ 
with the mean magnitude of the remaining   DLA-QSOs
with lower $\Delta(r-z)$. The most reddened cases are slightly fainter
on the average,
$\langle r \rangle = 19.24 \pm 0.08$,   
than the remaining cases, $\langle r \rangle = 18.98 \pm 0.04$.
A similar computation for the $(g-z)$ color yields
$\langle r \rangle = 19.11 \pm 0.09$ for the 30 most reddened cases
and $\langle r \rangle = 18.96 \pm 0.05$ for the remaining cases.
The effect is modest, but is in line with the prediction that
quasars with higher reddening should be statistically fainter. 

\begin{table}[htdp]
\caption{Mean reddening of sub-samples with different strengths of the \ion{Si}{ii} 1526 line
detected at the redshift of the DLA system.  }
\begin{center}
\begin{tabular}{| c | c | cc |cc | }
\hline 
 &   &      $W_{\ion{Si}{ii}} $& $< 1$ \AA\  &  $ W_{\ion{Si}{ii}} $& $\geq 1$ \AA\ \\
 \hline
color & Sample & $n$ &  $\langle E \rangle$ & $n$ &  $\langle E \rangle$ \\
      &        &     &   [10$^{-3}$ mag]    &     &  [10$^{-3}$ mag] \\
\hline
$(r-z) $ & 1& $151$ & $26 \pm 12$ & $29$ & $ 41 \pm 25$ \\
              & 2 & $150$ & $21 \pm 11$ & $29$ & $ 41 \pm 25$ \\
                & 3 & $149$ & $17 \pm 10$ & $29$ & $ 41 \pm 25$ \\
\hline               
$(g-z)$ & 1 & $142$ & $51 \pm 14$  & $29$ & $114 \pm 40$ \\
              & 2 & $141$ & $46 \pm 13$  & $29$ & $114 \pm 40$\\      
                & 3 & $140$ & $42 \pm 13$  & $29$ & $114 \pm 40$\\               
\hline 
\end{tabular}
\end{center}
\label{default}
\end{table}%

  \begin{figure*}
   \centering
 \includegraphics[width=8.5cm,angle=0]{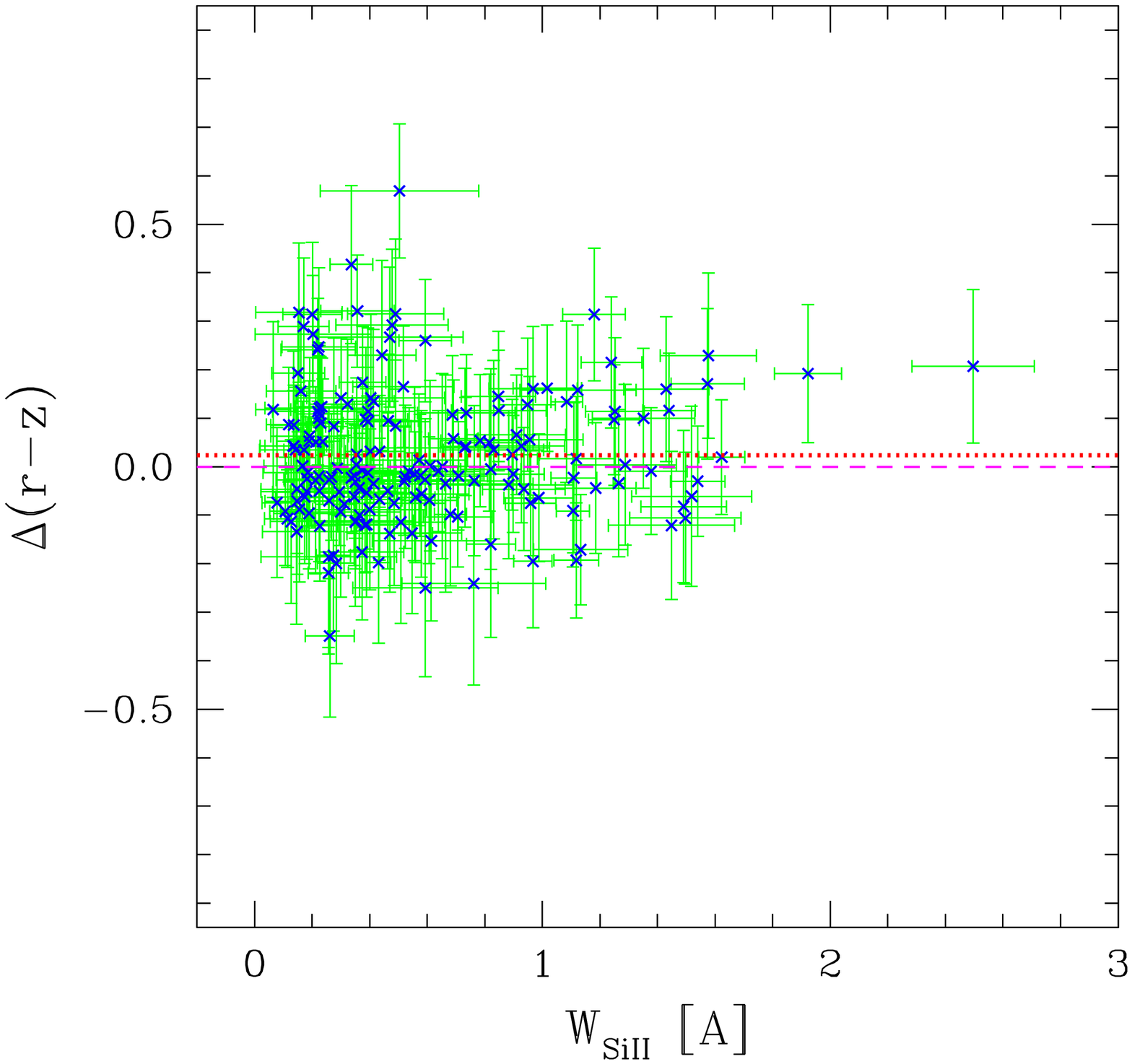}  
 \hskip 0.5cm
 \includegraphics[width=8.5cm,angle=0]{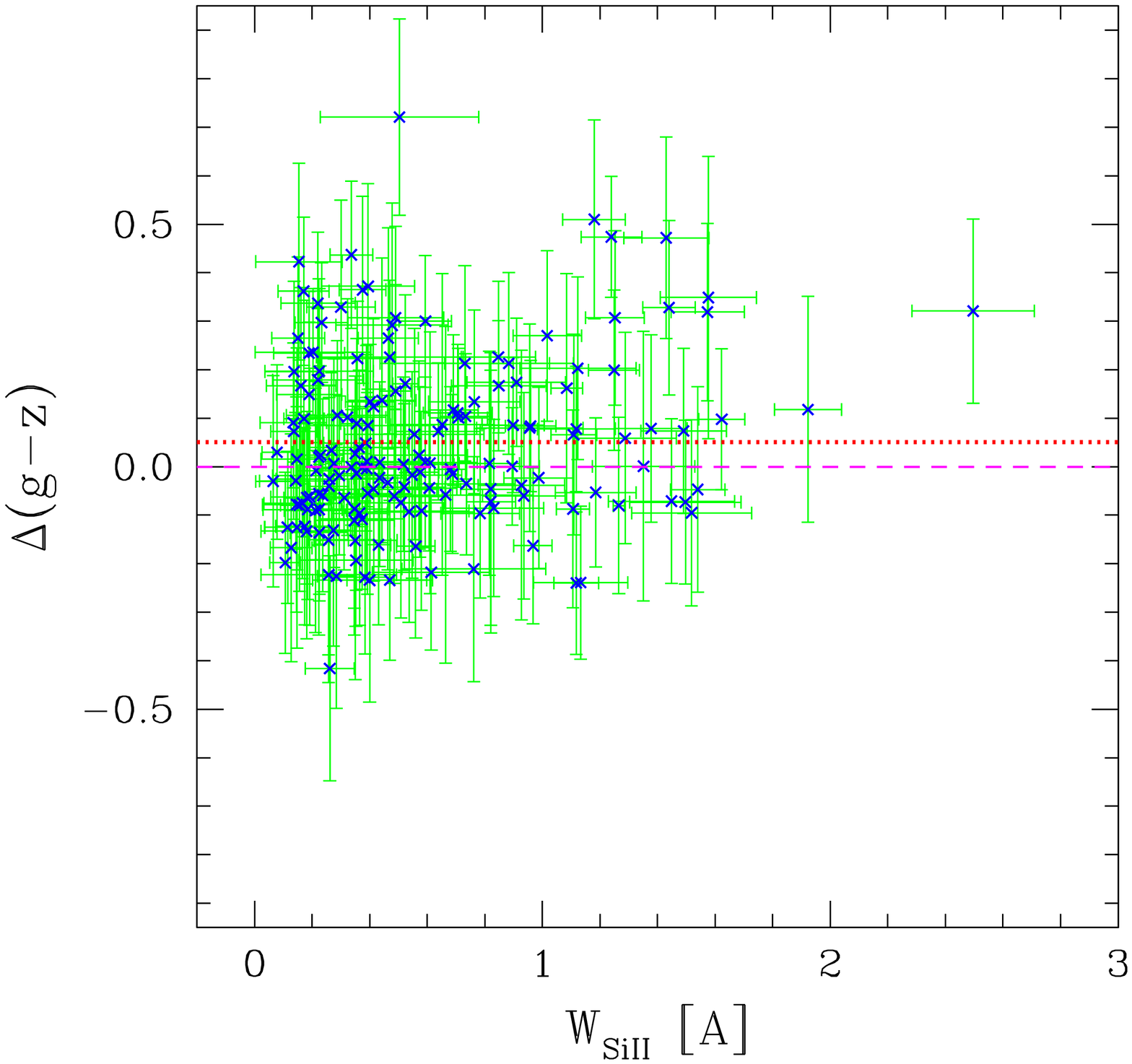} 
  \caption{ Deviations from the mean color, $\Delta(r-z)$ and $\Delta(g-z)$, versus  
   rest-frame equivalent width of \ion{Si}{ii} line 1526\AA, when detected,   for sample 2
    of Table 3.  Dotted lines: mean color excess of the same sample. 
  }
              \label{fig-Drz-WSi}%
    \end{figure*}

\subsection{Reddening versus metal column density}

If the reddening originates in dust embedded in the DLA system we expect
a trend between the color excess and the total column density of metals  
in the absorber.
The limited spectral resolution of SDSS spectra prevents the accurate measurement
of metal column densities, but is sufficient to measure the equivalent width of strong metal lines. 
To search for a possible trend between color excess and metals
we performed a study
of the \ion{Si}{ii} line at 1526\,\AA, which is
sufficiently strong to be detected in SDSS spectra
and lies      
redwards of the Ly\,$\alpha$ forest in most of the redshift range of interest. 
Since the 1526\,\AA\ line is saturated,
its rest-frame equivalent width, $W_\ion{Si}{ii}$, is not a good tracer
of the metal column density.
Nevertheless, an empirical correlation between $W_\ion{Si}{ii}$ and metallicity has been
recently found for DLA-QSOs, in the form
[M/H] $\simeq -0.92 + 1.41 \, \log W_\ion{Si}{ii}$ (Prochaska et al. 2007). 
This correlation is probably the consequence of the     mass-metallicity and   mass-kinematics relations in galaxies. 
Galaxies with higher masses (and metallicities) have larger kinematical motions, which result in larger
spread (and equivalent width) of saturated metal lines such as the \ion{Si}{ii} line at 1526\,\AA.

To estimate $W_\ion{Si}{ii}$ we used an automated algorithm
  which fits a Gaussian profile to the strongest absorption feature (if any)
  at the predicted wavelength, $\lambda_{\ion{Si}{ii}} = (1+z_a) \, 1526.707$\,\AA,
  provided this line occurs redwards of the Ly\,$\alpha$ forest.
  Each fit was visually inspected and the line-strength of \ion{Si}{ii}~1526
  was compared against other metal-lines observed redwards of the Ly$\alpha$
  forest.  In a few cases, we rejected lines because of obvious blends with
  coincident absorption lines from systems at unrelated redshifts.   As a result
  of this analysis, we measured an equivalent width 
  in 180 DLA-QSOs of our sample.
In Fig. \ref{fig-Drz-WSi} we plot
the deviations from the mean colors  $\Delta(r-z)$ and $\Delta(g-z)$
versus
the rest-frame equivalent width $W_\ion{Si}{ii}$
measured by the automated algorithm.
No clear trend is present, possibly because
most of the data are clumped at 
low values of equivalent width,
where we expect  a modest reddening.
The interval  
$W_\ion{Si}{ii} > 1 \, \AA$, where the reddening is expected
to be stronger, is much less populated, but its size is sufficiently
large to  compute the mean value.
We therefore splitted the data in two subsets of
``weak'' and ``strong'' \ion{Si}{ii} absorbers by using
$W_\ion{Si}{ii} = 1 \,\AA$ as a threshold value.
The mean values of reddening derived for these subsets are shown in Table 4
for the three absorption samples defined in Table 3.
One can see that 
the  mean  reddening tends to be higher
in strong \ion{Si}{ii}  absorbers     than in  
  weak \ion{Si}{ii} absorbers, as expected
  if the dust is embedded in DLA systems.
The difference in mean reddening between the two sub-samples is
marginal in the $(r-z)$ color, but more evident in the $(g-z)$ color.

To investigate the physical relation between dust and metals
we converted the color deviations $\Delta$ into a mean extinction 
and the \ion{Si}{ii} equivalent widths into a metallicity.  
The rest-frame extinction, $A_V$, is expected to scale with the metal column density 
$\mathrm{(M/H)} \times N(\ion{H}{i})$, where (M/H) is the total abundance by number
(gas plus dust) of the reference element M  
(e.g. Vladilo et al. 2006).

To derive the   extinction we 
used the expression 
$
  A_V    = 
    E(\lambda^\mathrm{obs}_y-\lambda^\mathrm{obs}_x)   \, / \, 
 \delta  \xi \, (z_a ; \lambda^\mathrm{obs}_x, \lambda^\mathrm{obs}_y)
    ,
$
 obtained from Eq. (\ref{Erz_efficiency}).
 The true color excess $E_i$ of the individual quasar is unknown,
 but we can take $E_i  \simeq \Delta_i$
 since we average the above expression for a large sample  
 to obtain  $\langle A_V \rangle$ (see Section 3.1).
 An SMC extinction curve $\xi (\lambda)$ was adopted in this conversion.

To derive the   metal column density we used
  the  empirical relation 
[M/H] $\simeq -0.92 + 1.41 \, \log W_\ion{Si}{ii}$ (Prochaska et al. 2007) and
 then averaged the individual values of metal column density 
$10^{\mathrm{[M/H]} + \log N(\ion{H}{i})}$.

\begin{figure}
   \centering
 \includegraphics[width=8.5cm,angle=0]{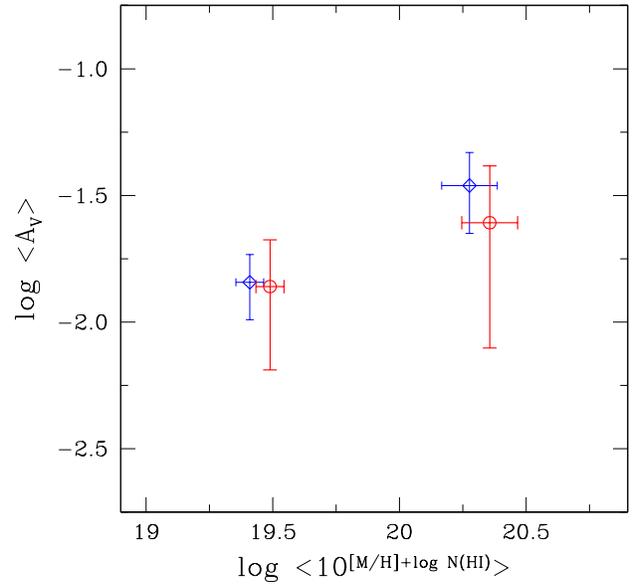} 
  \caption{ 
  Mean values of rest-frame extinction  and metal column density 
  of the two subsets of DLA-QSOs
  with different strengths of the \ion{Si}{ii} line 1526\,\AA\
  shown in Table 4 (absorption sample 2).
  The extinction values derived from the $(g-z)$ and $(r-z)$ color deviations
  are plotted with diamonds and circles, respectively. 
    The mean values of metal column density of the two subsets are 
  the same and have been slightly shifted along the horizontal
  direction for clarity. See Section 5.3.
  }
              \label{fig-AVMetcol}%
    \end{figure}

In Fig. \ref{fig-AVMetcol} we plot the results of these computations
obtained for the subsets of weak and strong \ion{Si}{ii} absorbers
of Table 4 (sample 2). 
The data are consistent with the existence of the trend
between metal column density and extinction  expected if 
the reddening and metals originate in the same medium.
Again, the evidence for this trend is very weak in $(r-z)$ and firmer in $(g-z)$.
Given the uncertainties in the above conversions and the limited statistics
it is not possible to reach firmer conclusions.

\begin{figure*}
   \centering
 \includegraphics[width=5.85cm,angle=0]{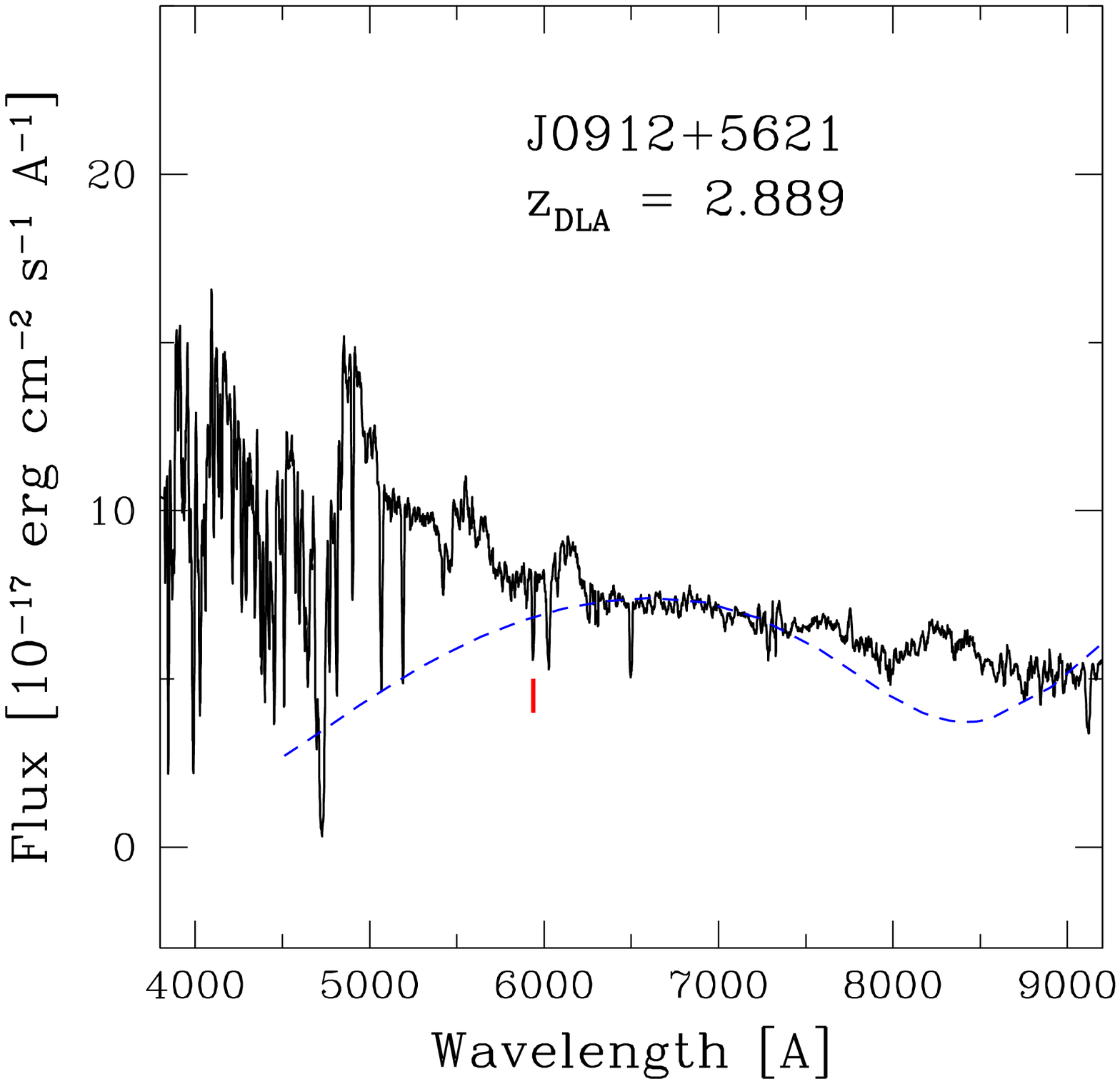}  
 \hskip 0.1 cm
 \includegraphics[width=5.85cm,angle=0]{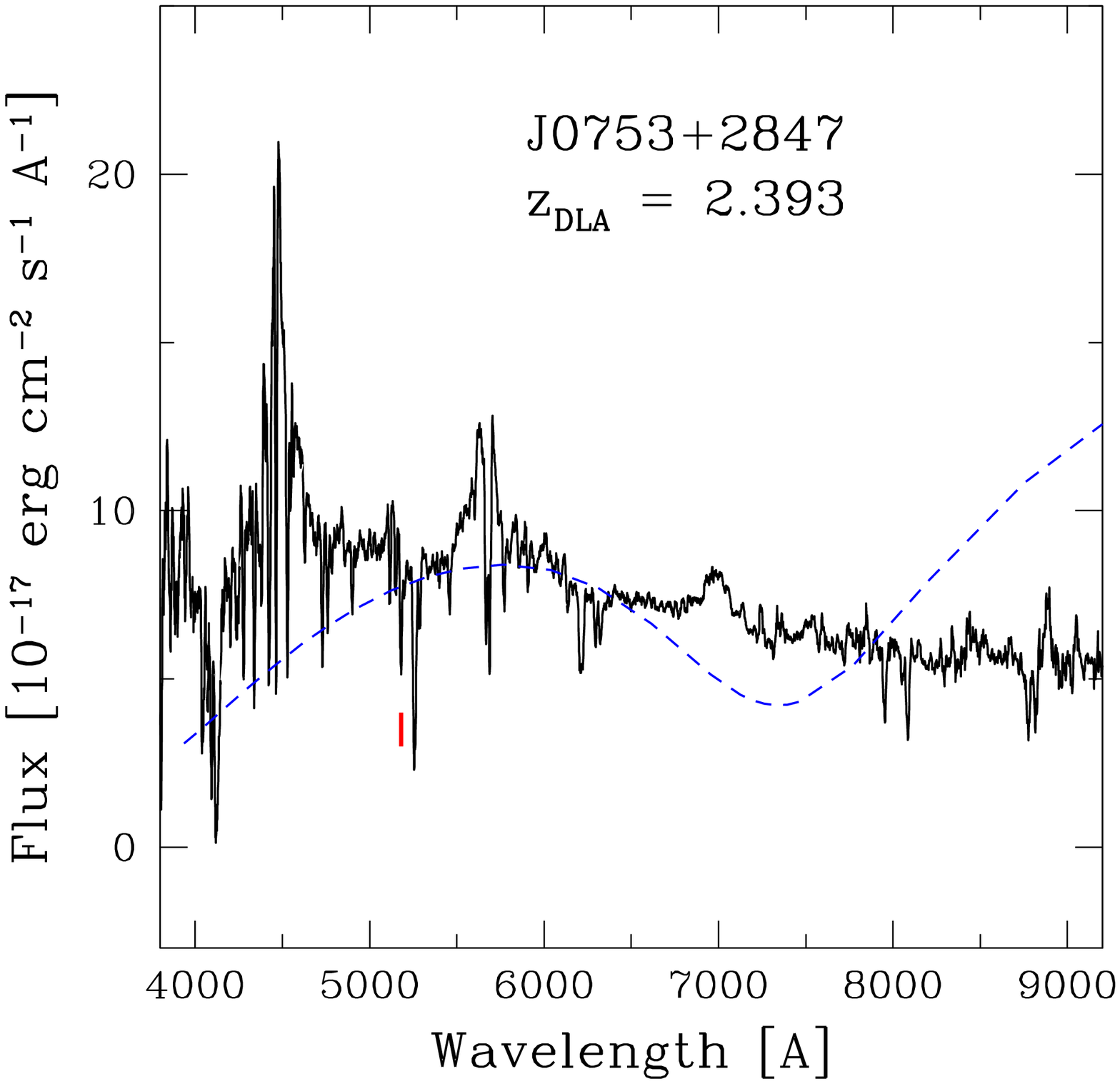} 
 \hskip 0.1 cm
 \includegraphics[width=5.85cm,angle=0]{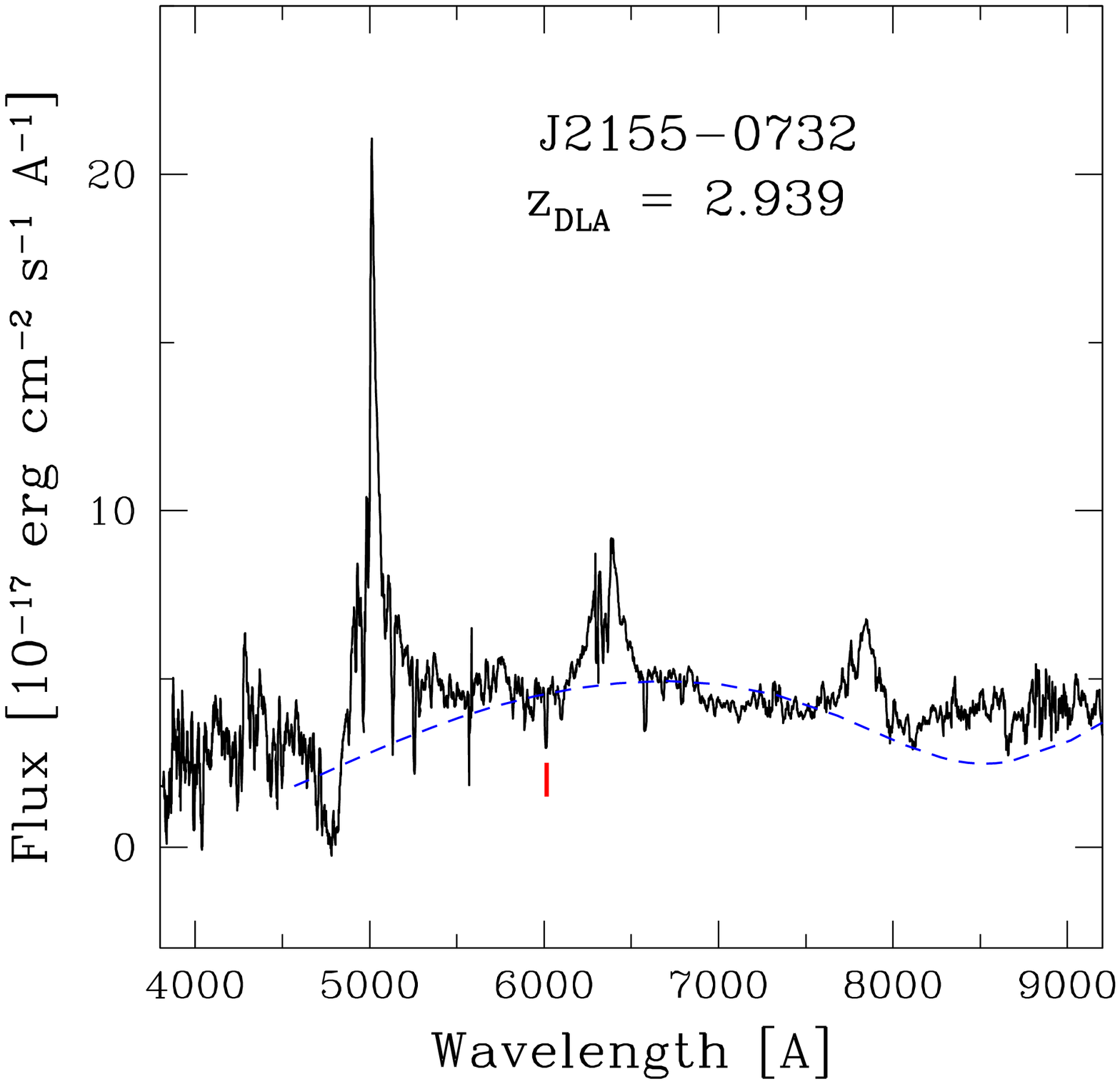} 
  \caption{ 
  Quasars of the absorption sample with color deviations $\Delta(g-z) > 0.4$ mag and strong \ion{Si}{ii} absorptions at $z_\mathrm{SiII} = z_\mathrm{DLA}$.  Dashed line: profile
  of the 2175\,\AA\ extinction bump plotted in a scale 
  $10^{-\xi[ \lambda / (1+z_\mathrm{DLA}) ]}$ (see Vladilo et al. 2006).
  }
              \label{fig-redquasars}%
    \end{figure*}

The quasars of the absorption sample with high color deviations
and strong \ion{Si}{II} absorptions are excellent candidates for detailed
follow up studies of dusty DLA systems based on high resolution spectroscopy. 
In Fig. \ref{fig-redquasars} we show, as an example, 
the spectra of the DLA-quasars with $\Delta(g-z) > 0.4$
and $W_\ion{Si}{ii} > 1 \,\AA$. 
 The typical metallicity of these systems is a factor 3 larger  than  the average metallicity of the absorption sample.

\subsection{Dust-to-gas ratio}

By adopting an extinction curve $\xi(\lambda)$ of  SMC-type 
we can  convert  the reddening measured in the observer's frame  into 
an extinction in the rest frame.
From this in turn we can measure, for the first time in studies of DLA systems,  
the dust-to-gas ratio in the same units used in local
interstellar studies, namely the ratio $A_V / N(\ion{H}{i})$
between the extinction in the $V$ band and the \ion{H}{i} column density.
From Eq.(\ref{Erz_efficiency})   
the individual dust-to-gas ratio is given by
\begin{equation}
{ A_V \over N(\ion{H}{i}) } = 
{   E(\lambda^\mathrm{obs}_y-\lambda^\mathrm{obs}_x)   \, / \, 
 \delta  \xi \, (z_a ; \lambda^\mathrm{obs}_x, \lambda^\mathrm{obs}_y)
    \over N(\ion{H}{i}) } ~.
 \end{equation}
Since we average this quantity for all the systems of our sample
we  take $E_i  \simeq \Delta_i$  (see Section 3.1).
The mean values obtained from this computation  are shown in Table 3.
We obtain
$\langle A_V / N(\ion{H}{i}) \rangle \simeq$  3 to 4 $\times 10^{-23}$ mag cm$^2$,
from  the measurements in the $(r-z)$ or $(g-z)$ colors, respectively.
By excluding the most reddened cases  
we obtain 
 $\langle A_V / N(\ion{H}{i}) \rangle \simeq$ 2.5 to 3.5 $10^{-23}$ mag cm$^2$.
%
For comparison, the typical dust-to-gas ratio of the Milky Way is
$A_V / N(\ion{H}{i}) \approx 5.3 \times 10^{-22}$ mag cm$^2$
(Bohlin et al. 1978).
These results indicate that the dust-to-gas ratio of DLA systems
is deficient by $\approx -1.25$  dex relative to that of the Milky Way.
This deficiency is consistent with the lower level of metallicity
of the DLA systems of our sample. This metallicity can be estimated from the
relation between [M/H] and $W_\ion{Si}{ii}$ (Prochaska et al. 2007). 
For the systems with measurements of $W_\ion{Si}{ii}$ we obtain
a mean value
$\langle \mathrm{[M/H]} \rangle \simeq -1.2$ dex (Table 5).
We conclude that the dust-to-gas ratio scales with the metallicity, 
as expected for dust embedded in the DLA galaxies.

\begin{table*}[htdp]
\caption{Mean metallicity and extinction per unit metal column density for the DLA-QSOs with \ion{Si}{ii} measurements.}
\begin{center}
\begin{tabular}{| c|cccc|cccc | }
\hline 
color & & & $(r-z)$   & & & & $(g-z)$     & \\
\hline
 Absorption 
 &   $n$   & ${\langle \mathrm{[M/H]} \rangle}^b$ & ${\langle \mathrm{[M/H]} \rangle}_w^c$ & $\langle s_V^\mathrm{Fe} \rangle^d$  
 &   $n$   & ${\langle \mathrm{[M/H]} \rangle}^b$ & ${\langle \mathrm{[M/H]} \rangle}_w^c$ & $\langle s_V^\mathrm{Fe} \rangle^d$ \\
  sample$^a$          &           &  &     & [10$^{-17}$ mag cm$^2$]          
   &           &   &   & [10$^{-17}$ mag cm$^2$] \\
\hline
1  & 180 & $-1.19 \pm 0.03$ & $-1.07$ & 14 $\pm$ 8 & 171 & $-1.18 \pm 0.03$ & $-1.07$ & 7.5 $\pm$ 4.3\\ 
2  & 179 & $-1.18 \pm 0.03$ & $-1.07$ & 12 $\pm$ 8 & 170 & $-1.18 \pm 0.03$ & $-1.06$ & 5.8 $\pm$ 4.0\\
3  & 178 & $-1.18 \pm 0.03$ & $-1.07$ & 11 $\pm$ 8 & 169 & $-1.17 \pm 0.03$ & $-1.06$ & 5.4 $\pm$ 4.0\\  
\hline 
\end{tabular}
\end{center}
\vskip 0.2cm
\noindent
$^a$ Absorption samples are defined in Table 3. Only cases with \ion{Si}{ii} line detected
at more than 1 $\sigma$ level are considered in this table. \\  
\noindent
$^b$ Mean metallicity $\langle \mathrm{[M/H]} \rangle \equiv \log \, \langle 10^\mathrm{[M/H]}\rangle$. 
The quoted error is the standard error of the mean.\\  
\noindent
$^c$ Mean \ion{H}{i} column-density weighted metallicity of Eq. (\ref{HIWeightedDef}). \\  
\noindent
$^d$ Mean extinction per unit dust column density of iron (Section 5.5).
The quoted error is the standard error of the mean.\\  
\label{cucu}
\end{table*}%

\subsection{Dust-to-metal ratio}

The present measurements of reddening and metallicity can be combined to
estimate the mean extinction per unit metal column density in the DLA systems of our sample. 
We start from  the  relation
\begin{equation}
A_V = \langle s_V^\mathrm{Fe}\rangle \times f_\mathrm{Fe} \times \mathrm{(Fe/H)} \times N(\ion{H}{i}) ~,
\label{AVsV}
\end{equation} 
where $\langle s_V^\mathrm{Fe}\rangle$ is the mean extinction per unit dust column density of iron
and $f_\mathrm{Fe}$ is the fraction of iron in dust form (Vladilo et al. 2006). 
Previous measurements of
 $\langle s_V^\mathrm{Fe}\rangle$  have yielded a typical value\footnote
{
Mean value of the most accurate
measurements of $A_V$ and dust column density $\widehat{N}_\mathrm{Fe}$
for the  DLA systems in Table 3 by Vladilo et al. (2006); 
the same value   matches   well
the interstellar trend between $\widehat{N}_\mathrm{Fe}$ and $A_V$  
(see Fig. 4 in the same paper). 
}
 $\langle s_V^\mathrm{Fe}\rangle \approx 3 \times 10^{-17}$ mag cm$^2$, 
roughly constant within a factor of two,
in different types of \ion{H}{i} interstellar regions, including 
metal absorption systems and a few
DLA systems   at $z_a \lsim 2$. 
The present data set allows us to investigate whether or not this parameter 
has a similar value also in the general population of
DLA systems at $z_a \gsim 2$.
 
By combining the Eqs. (\ref{AVsV}) and (\ref{Erz_efficiency}) we obtain    
\begin{equation}
 \langle s_V^\mathrm{Fe}\rangle =
{ E(\lambda^\mathrm{obs}_y-\lambda^\mathrm{obs}_x)   \, / \, 
 \delta  \xi \, (z_a ; \lambda^\mathrm{obs}_x, \lambda^\mathrm{obs}_y)
  \over 
  f_\mathrm{Fe} \times  (\mathrm{Fe/H})  \times N(\ion{H}{i})   }  ~.
\end{equation}
As in the previous section, we  
take $E_i  \simeq \Delta_i$ because
we average this expression for all the systems of our sample. 
We adopt an  SMC extinction curve, 
assume (Fe/H) $=$ (M/H) and take $f_\mathrm{Fe} \simeq 0.65$,
a value appropriate for DLA systems at the redshift of our sample    
(Vladilo 2004).
The resulting values of $\langle s_V^\mathrm{Fe}\rangle$, shown in Table 5,
are broadly consistent with the value 
$\langle s_V^\mathrm{Fe}\rangle \approx 3 \times 10^{-17}$ mag cm$^2$
inferred from previous investigations\footnote{
The recent result
$\langle E(B-V) \rangle / \langle N_\mathrm{Zn} \rangle \simeq 4.1 \times 10^{-15}$mag cm$^2$,  found 
by Menard et al. (2007) in  \ion{Mg}{ii} systems, yields the same value of $\langle s_V^\mathrm{Fe}\rangle$, assuming that   
the Zn/Fe ratio is solar and
$f_\mathrm{Fe} \simeq 0.65$.
} 
at lower redshift.
This is true, in particular, for 
our best estimate  obtained from the $(g-z)$ color deviations.
Considering the uncertainties in the values of $f_\mathrm{Fe}$ and (Fe/H),
this broad agreement is rather remarkable and suggests that the
extinction properties of the dust are similar to those of low-redshift dust.
A firmer conclusion must await more accurate determinations of the metallicities
and depletions of the present sample.

  \begin{figure*}
   \centering
 \includegraphics[width=8.5cm,angle=0]{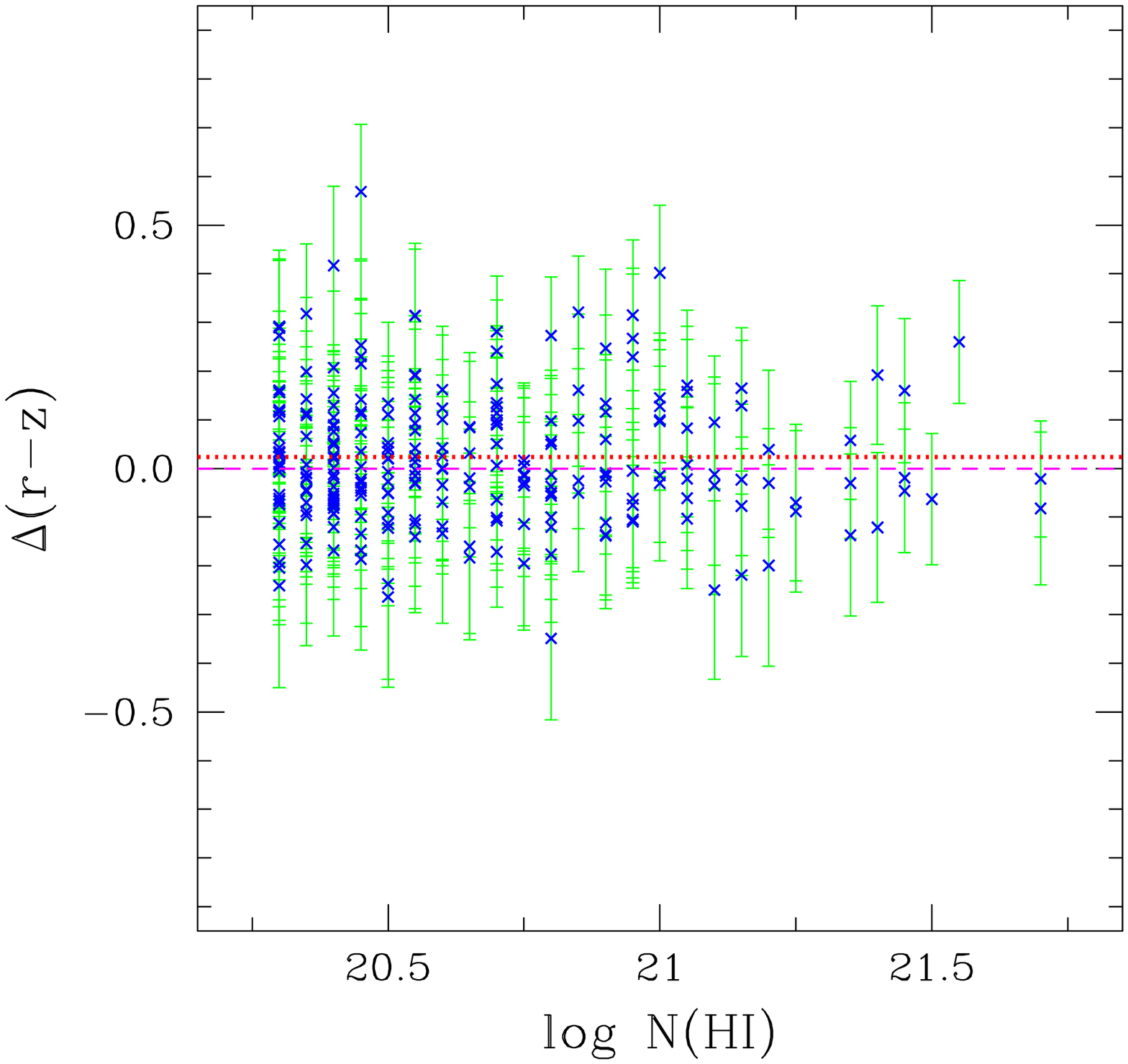}  
 \hskip 0.5cm
 \includegraphics[width=8.5cm,angle=0]{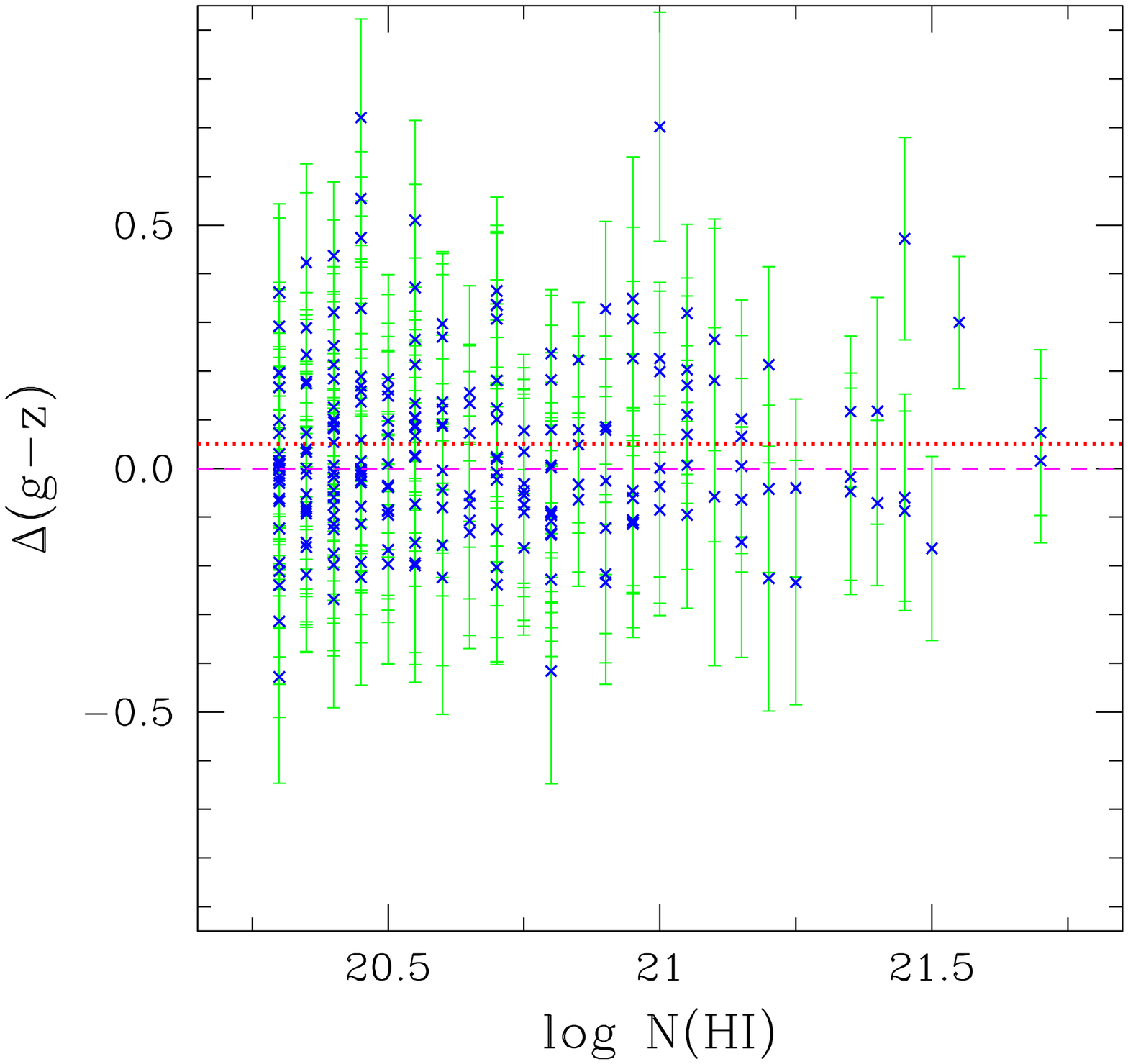} 
  \caption{ Deviations from the mean color, $\Delta(r-z)$ and $\Delta(g-z)$, versus \ion{H}{i} 
  column density for sample 2  of Table 3.
    Dotted lines: mean color excess of the same sample. 
  }
              \label{fig-Drz-HI}%
    \end{figure*}

\subsection{Reddening and \ion{H}{i} column density }

In Fig. \ref{fig-Drz-HI} we plot the color deviations 
$\Delta(r-z)$ and $\Delta(g-z)$ versus \ion{H}{i} column density.
A linear correlation analysis    
of $\Delta(r-z)$
and  $\Delta(g-z)$ versus  $\log N$(\ion{H}{i}) 
for the complete absorption sample
yields correlation coefficients $r=-0.01$ and $r=+0.01$, respectively.
The comparison of sub-samples with low and high values of $N$(\ion{H}{i})
does not indicate the existence of a trend with reddening.
%
Most DLA-QSOs are concentrated at low 
\ion{H}{i} column densities as a consequence of the
well-known decrease of the number of DLA systems
with increasing $N(\ion{H}{i})$ (Wolfe et al. 2005; Prochaska et al. 2005).
Some of the most reddened quasars are concentrated 
at low \ion{H}{i} column densities,  at odd with
the expectation that large reddening  is associated with 
large column density of neutral gas.
The paucity of highly reddened cases at $N(\ion{H}{i}) > 10^{21}$ atoms cm$^{-2}$
could be the consequence of low-number statistics:
  systems with high \ion{H}{i} column density are rare and
the present statistics might be insufficient to detect  cases of high reddening.
It is worth noticing that lines of sight of high metallicity are 
generally not observed in DLA systems with
$N(\ion{H}{i}) > 10^{21}$ atoms cm$^{-2}$ (Boiss\'e et al. 1998).
If this is the case also for the present sample,
this fact could conspire to make rare the systems of high reddening 
at $N(\ion{H}{i}) > 10^{21}$ atoms cm$^{-2}$. 

\subsection{The \ion{H}{i} column-density weighted metallicity}

The metal abundances of DLA systems are often used to estimate
the mean cosmic metallicity of the neutral gas at high redshift.
In this type of estimate the metallicity of each system 
must be weighted by its \ion{H}{i} column density to estimate
a mean cosmic value
\begin{equation}
\langle \mathrm{[M/H]}   \rangle_w = 
{\sum_i \, (\mathrm{M/H})_i \times N_i(\ion{H}{i}) \over \sum_i N_i(\ion{H}{i})} 
\label{HIWeightedDef}
\end{equation}
 (Lanzetta et al. 1995). 
We estimated the above expression
by using the indirect measurement of metallicity obtained from
the equivalent width of the \ion{Si}{ii} line. The results are shown in Table 5.
We obtain
$ \langle \mathrm{[M/H]}   \rangle_w \simeq -1.1$ dex. 
This value is somewhat higher than
the mean \ion{H}{i}-weighted  metallicity of DLA
systems measured in high-resolution surveys 
 at redshift $z_a \approx 3$,
which is   $\simeq -1.5$   dex   
(Prochaska et al. 2001; Kulkarni et al. 2005). 
The present survey of low-resolution SDSS quasar  spectra
is characterized by a higher magnitude limit 
since it includes quasars as faint as $m_r = 20.2$. 
The higher value of metallicity of the present SDSS sample 
suggests that the mean metallicity of DLAs
may increase with the limiting magnitude of the survey.
An effect of this type is predicted as a consequence of 
the dust extinction bias (Ellison et al. 2001, Vladilo \& P\`eroux 2005),
but the empirical evidence for its existence is still lacking (Akerman et al. 2005).
An accurate determination of the metallicities of the present absorption sample
will offer us a way to test the dependence of the mean
metallicity on  the limiting magnitude of DLA-QSO surveys.

  \begin{figure*}
   \centering
 \includegraphics[width=8.5cm,angle=0]{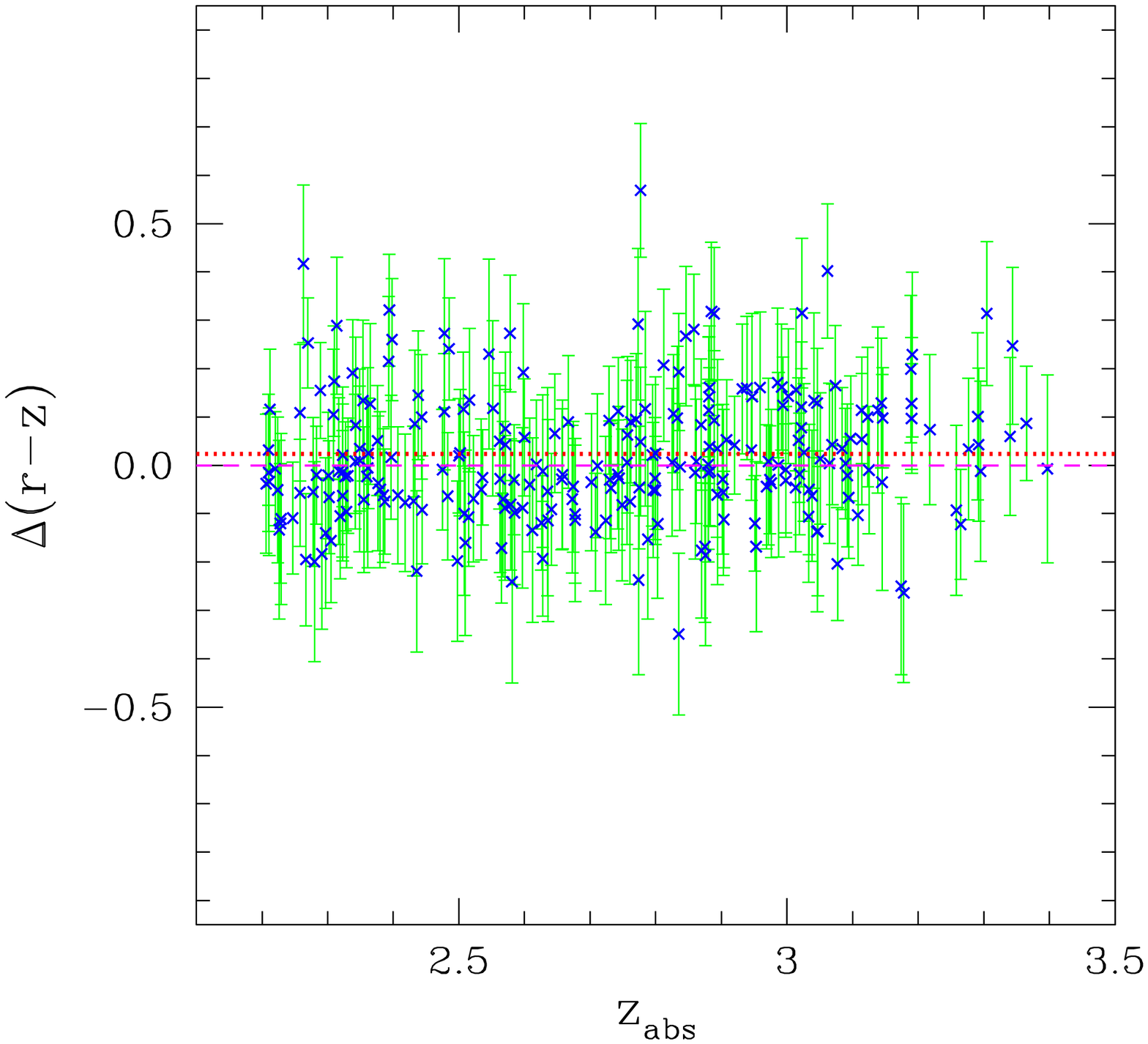}  
 \hskip 0.5cm
 \includegraphics[width=8.5cm,angle=0]{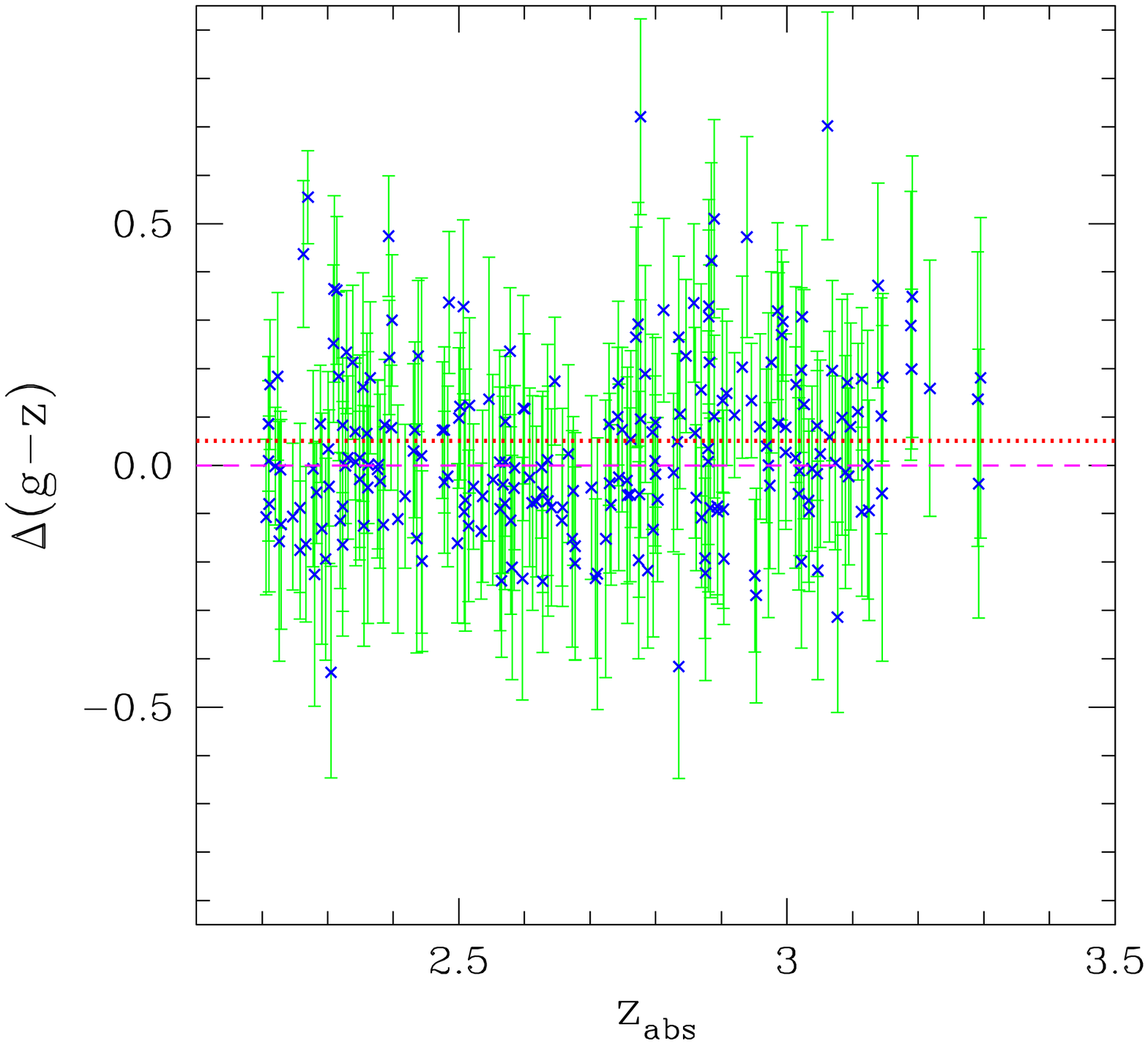} 
  \caption{ Deviations from the mean color, $\Delta(r-z)$ and $\Delta(g-z)$, 
  versus absorption redshift for sample 2 of Table 3.
    Dotted lines: mean color excess of the same sample. 
  }
              \label{fig-Drz-zab}%
    \end{figure*}

\subsection{Reddening versus DLA redshift}

In Fig. \ref{fig-Drz-zab} we plot the color deviations $\Delta(r-z)$
and $\Delta(g-z)$ versus absorption redshift $z_a$.
%
The data
are spread more or less uniformly
in the interval of absorption redshift considered.
A linear correlation analysis 
of $\Delta(r-z)$
and  $\Delta(g-z)$ versus $z_a$ 
for the complete absorption sample yields 
correlation coefficients $r=+0.13$ and $r=+0.06$, respectively.
%
The lack of a trend with redshift implies
that the color excess that we derive is   sufficiently
representative of the mean reddening of
DLA systems over the full interval   $2.2 \lsim z_a \lsim 3.5$   covered by our sample.
%
The lack of redshift evolution is, to some extent, surprising.
Two types of evolutionary effects 
are expected to occur if the reddening originates in DLA dust.
On the one hand, a rise of the reddening
with increasing $z_a$ is expected due the UV rise of the extinction curve.
For an SMC extinction curve this effect is roughly   a factor of 2 in $E(r-z)$
in the redshift range of interest, as shown in  Fig. \ref{fig-NErz-za}.
On the other hand, a decrease of the reddening
with increasing $z_a$ is expected as a consequence of cosmic chemical evolution
if the dust scales with the metals. Also this effect is expected to be approximately a factor of 2 
in the redshift interval of interest,  where the mean   metallicity decreases by
$\simeq 0.3$ dex per unit redshift interval (Prochaska et al. 2000).
The lack of a trend could be the consequence of the opposite behaviour
of two evolutionary effects, which tend to cancel each other.  


\subsection{Comparison with previous work}

To perform a homogeneous comparison
with the work  of  Murphy \& Liske (2004; ML04) we applied our procedure
to their original list of quasars\footnote{
See web address  
http://www.blackwellpublishing.com/products
/journals/suppmat/MNR/MNR8374/mnr8374TableS1.txt.
}
extracted from the second data release (DR2).
The list includes 70 DLA-quasars and 1396 non-DLA quasars
in spectra with SNR $\geq 3$ as defined in ML04.
To apply our procedure we selected  
the 56 DLA-quasars and 1074 non-DLA quasars with  $z_e \leq 3.5$.
The non-DLA quasars were used to build the control pool.
From the application of our method to these lists we obtain
the weighted mean values
$\langle E(r-z) \rangle = 3 \, (\pm 22) \times 10^{-3}$ mag  and,
for the subset of 52 quasars  with $z_e \leq 3.4$,  
  ${\langle E(g-z) \rangle}  = 39 \, (\pm 36) \times 10^{-3}$ mag.
The lack of detection of the reddening is due to the small size of the sample.
By converting these results to rest-frame color excess with an SMC
extinction curve  we obtain a $3 \sigma$ upper limit $E(B-V) < 0.02$ mag from both color indices.
This limit is equal to that obtained by ML04 from their analysis
of the spectral index distribution of the quasar spectra.
The photometric method presented here
and the spectral index method adopted by ML04  yield therefore consistent results.

%
%

\section{Summary and conclusions}

We have used the spectroscopic and photometric database of the 5th data release of the  SDSS 
to measure the reddening of quasars
with intervening Damped Ly $\alpha$ systems at 
high redshift ($z_\mathrm{DLA} > 2.2$). 
The spectroscopic database was used
to distinguish quasars with and without intervening absorption systems. 
Only good quality spectra, with typical
signal-to-noise ratio SNR $\gsim 4$ in the spectral regions of interest,
were considered in the analysis.
We built up an ``absorption sample'' of 248 quasars with a single DLA system
in the redshift interval $2.2 < z_a \leq 3.5$ and without metal systems
at $z_\mathrm{metals} \neq z_\mathrm{DLA}$. 
We then selected about 2 thousand control quasars without DLA systems
or metal systems. 
%
%
The large SDSS photometric database, employed
for the first time in  the DLA reddening analysis,
was used to measure the colors 
of individual quasars of the absorption and control samples.   

The analysis was performed on the $(r-z)$ and $(g-z)$ color indices. 
The color index $(r-z)$ is uncontaminated by Ly  $\alpha$ absorption
in the redshift interval considered.
The color $(g-z)$ offers a higher leverage for reddening detection,
but the   $g$ bandpass is contaminated by Ly $\alpha$ absorption.
To derive the $(g-z)$ colors we
corrected the $g$ magnitude
for flux suppresion due to the presence of the damped Ly  $\alpha$ absorption
 in the spectra.
We obtain   a mean color excess 
$\langle E(r-z) \rangle = 27 \, (\pm 9) \times 10^{-3}$ mag 
and  ${\langle E(g-z) \rangle}  = 54 \, (\pm 12) \times 10^{-3}$ mag.
The quoted statistical error is not significantly affected by the presence
correlated errors. 
The detection is confirmed by the analysis of 10,000 bootstrap
samples originated from the absorption sample. 
The reddening is detected at $\approx 3 \, \sigma$ c.l. also when
the most reddened DLA-QSOs of the sample are rejected, 
assuming that they are spurious.

The most natural hypothesis to explain the observed reddening 
is dust along the line of sight. 
The ratio $\langle E(g-z) \rangle/\langle E(r-z) \rangle = 2.2 \pm 0.9$ 
that we measure lends support to this hypothesis, since this
value is in agreement with that predicted
for dust with SMC-type extinction curve at the absorption redshift of our sample.
Also the study of the reddening versus quasars magnitude 
is consistent with the dust hypothesis: the most reddened quasars
are, on the average, slightly fainter than the others, in line with the expectation that 
  extinction and  reddening should increase together  if the reddening is due to dust.

Dust in the quasar environment 
 or in low-redshift interlopers
is  not a viable explanation for the observed reddening
 since it would affect in the same way, on the average, 
the quasars of the absorption sample and those of the control sample.
Quasars with anomalous colors would not be able to produce accidentally 
the measured reddening  
(Section 4.1).

The above arguments indicate that the reddening is due to
dust embedded in the  intervening DLA systems. 
This conclusion is consistent with two other
studies performed in the present work.
One is the comparison of  the color excess with the equivalent width 
of the \ion{Si}{ii} 1526\AA\ line at the redshift of the DLA system:
strong \ion{Si}{ii} absorbers  ($W_\ion{Si}{ii} \geq 1 \, \AA$) show
a slightly higher reddening, on the average,
than weak \ion{Si}{ii} absorbers  ($W_\ion{Si}{ii} < 1 \, \AA$),
in line with the expectation that the amount of dust and metals 
should increase together inside the DLA systems.
Also the study of the mean dust-to-gas ratio, obtained by converting the observed
reddening into rest-frame extinction, yields a similar conclusion. 
We find  
$\langle A_V / N(\ion{H}{i}) \rangle \approx$  2 to 4 $\times 10^{-23}$ mag cm$^2$,
a value $\approx -1.25$ lower
than that of the Milky Way, 
  consistent with the lower level of metallicity  
of DLA systems of our sample
estimated indirectly from  $W_\ion{Si}{ii}$. 
Also this result is  in line with
the expectation that metals and dust should  increase together.

The conversion of our reddening measurement
to rest-frame  color excess 
by means of an SMC extinction curve yields 
$\langle E(B-V) \rangle \simeq 0.005/0.007$ mag.
This value fits the trend of $\langle E(B-V) \rangle$ 
versus redshift in quasar absorbers extrapolated
from measurements of \ion{Mg}{ii} systems at lower redshift (Menard et al. 2007; Fig. 10). 
  
The mean color excess that we have derived is
representative of the mean reddening of DLA systems at  $z_a \approx 2.7$
in SDSS QSOs with limiting magnitude $r \simeq 20.2$.  
The magnitude limit results from the   
condition SNR $\gsim 4$, required to distinguish the presence of
absorption systems in the quasar spectra.   
Since highly reddened quasars are more likely to be found  at  
fainter magnitudes, a larger reddening could be obtained from deeper surveys. 

Follow-up, direct measurements of metallicity of the DLA systems of the present  sample
are required to test the indirect estimates based on $W_\ion{Si}{ii}$,
  to obtain an accurate measurement of the extinction per unit metal column density, 
which seems to be similar to that found at lower redshift,
and to assess
the importance of the extinction bias in studies of DLA-QSOs.

\begin{acknowledgements}
       We thank Gabriel Prochter for providing an updated list   
      of \ion{Mg}{ii} absorptions in SDSS DR5 quasar spectra.
      G.V. thanks Matteo Viel, Sergei Levshakov and Carlo Morossi for helpful discussions
      on the analysis of the data. 
      J.X.P. is partially supported by an NSF CAREER grant (AST-0548180).
       We thank an anonymous referee for comments that improved this paper. 
\end{acknowledgements}


\appendix 
 
 \section{Derivation of Eq. (\ref{equality}).}

By comparing the definitions 
$E=C-C_0$  and 
$\Delta = C - \,  \langle C^u  \rangle$ given in Section 3.1,
  we have
\begin{equation}
\Delta
 = E \, + \, \Delta_{\, 0}   \,  ,
\label{DELTA2}
\end{equation}
where
$\Delta_{\, 0} =  C_{0 } -   \langle C^u  \rangle$.
%
The quantities $E$ and $\Delta_{\, 0 } $ are physically independent because
the color excess $E$ depends on the physical properties of the
absorber, while
    $\Delta_{\, 0 } $  on  the intrinsic colors of the quasars.
Since $E$ and $\Delta_{\, 0 } $ are   independent variables,
  the frequency distribution of their sum 
  equals the convolution product
of their frequency distributions, i.e.
$f_\Delta = f_E \otimes f_{\Delta_0} $.
%
From this result it follows that
\begin{equation}
 \langle \Delta \rangle  = \langle E \rangle  ~
 \label{equality2}
\end{equation}
 for any arbitrary distributions $f_E$ and $f_\Delta$, 
provided   $\langle\Delta_{\, 0 }\rangle=0$.  
This last condition  is valid 
for a large sample of reddened quasars with  same properties since
in this case the unknown unreddened color $C_{0 }$ belongs, by construction,
to the same population  of  the unreddened colors of the control quasars $C^u$.

If we have a large sample
of reddened quasars with varying properties $p_i$, then  the   condition 
$\langle \Delta_{\, 0 } \rangle=0$
is still   satisfied provided   the distributions of
unreddened colors  {\em around their mean value}
are the same for different subsets of control quasars.
The validity of 
the condition $\langle\Delta_{\, 0 } \rangle=0$ can be tested from
the analysis of quasars of the  twin control samples
(Section 4.1).  
Each control sample is built with unreddened QSOs and therefore its
  color deviations 
are  
$\Delta_{\, 0 } =  C_{0 } -   \langle C^u  \rangle$. 
By measuring the mean value of these   deviations 
we can   test  if     $\langle \Delta_{\, 0 } \rangle=0$.
From Fig. \ref{fig-Twins-distr} one can see that 
the mean values of the color excess of the  control samples are peaked at zero,
implying that the above condition is generally satisfied.

\end{document}